\documentstyle[12pt,epsf]{article}

\newlength{\dinwidth}
\newlength{\dinmargin}
\def\funp{{I\!\!P}}
\def\lapproxeq{\lower .7ex\hbox{$\;\stackrel{\textstyle<}{\sim}\;$}}
\def\gapproxeq{\lower .7ex\hbox{$\;\stackrel{\textstyle>}{\sim}\;$}}
\begin{document}
\titlepage
\begin{flushright}
DTP/96/50  \\
hep-ph/9606443 \\
June 1996 \\
\end{flushright}

\begin{center}
\vspace*{2cm}
{\Large \bf Diffractive open charm production at HERA} \\
\vspace*{1cm}
E.\ M.\ Levin$^{a,b,c}$, A.\ D.\ Martin$^a$, M.\ G.\ Ryskin$^{a,c}$ and 
T.\ Teubner$^a$, \\
\end{center}
\vspace*{0.5cm}

\indent $^a$ Department of Physics, University of Durham,
Durham, DH1 3LE, England.

\indent $^b$ LAFEX, Centro Brasileiro de Pesquisas Fisicas,
22290-180, Rio de Janeiro, Brazil.

\indent $^c$ Petersburg Nuclear Physics Institute, 188350 Gatchina, 
St.\ Petersburg, Russia.

\vspace*{2cm}
\begin{abstract}
We use perturbative QCD to calculate the cross sections
$\sigma^{L,T}$ for the diffractive production of open charm
$(c\overline{c})$ from longitudinally and transversely polarised
photons (of virtuality $Q^2$) incident at high energy
$(\sqrt{s})$ on a proton target.  We study both the $Q^2$ and
$M^2$ dependence of the cross sections, where $M$ is the invariant
mass of the $c\overline{c}$ pair.  Surprisingly, the result for
$\sigma^T$, as well as for $\sigma^L$, is perturbatively stable. 
We estimate higher-order corrections and find a sizeable enhancement 
of the cross sections.  The cross sections depend on the {\it square} of 
the gluon density $g (x, K^2)$, and we show that the observation of open 
charm at the HERA electron-proton collider can act as a sensitive probe of
the gluon distribution for $x = (Q^2 + M^2)/s$ and scale $K^2 =
(m_c^2 + \langle k_T^2 \rangle) (1 + Q^2/M^2)$ where the average
quark transverse momentum squared $\langle k_T^2 \rangle \sim
m_c^2$.  As compared to diffractive $J/\psi$ production, open
charm has the advantage that it is independent of the
non-perturbative ambiguities arising from the $J/\psi$ wave
function.  We estimate the fraction of diffractive events that arise 
from $c\overline{c}$ production.
\end{abstract}

\newpage
\noindent {\large \bf 1.  Introduction} \\

The recent observation of high energy diffractive $J/\psi$ photo-
and electroproduction, $\gamma^{(*)} p \rightarrow J/\psi \: p$, at
HERA \cite{jps} has attracted a lot of interest.  The principle
reason is
that the presence of the \lq\lq large" scale $M_\psi^2 + Q^2$
makes the process amenable to perturbative QCD, even for
photoproduction $(Q^2 = 0)$.  ($M_\psi$ is the mass of the
$J/\psi$ vector meson.)  Indeed for sufficiently high $\gamma p$
centre-of-mass energy $W$ the cross section for this, essentially
elastic, process can be expressed in terms of the {\it square} of
the gluon density.  Thus, in principle, it seems to offer a
particularly sensitive prove of the gluon distribution at small
$x$.

\begin{figure}[htb]
\begin{center}
\leavevmode
\epsfxsize=14.cm
\epsffile[100 360 540 570]{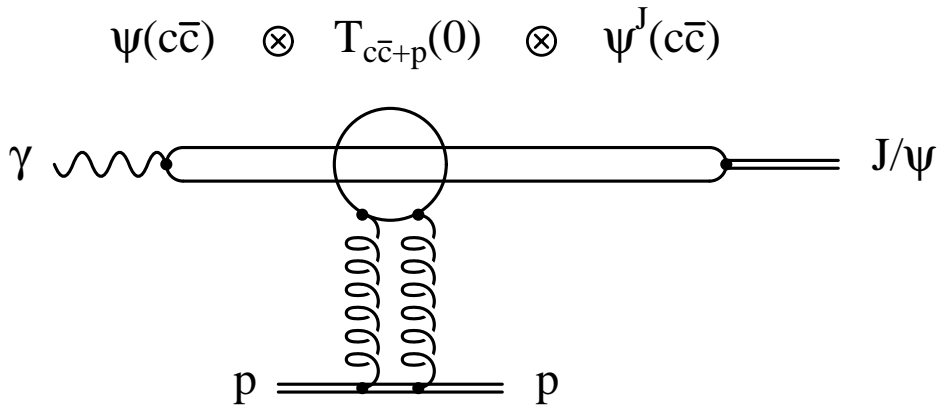}
\vskip -8mm
\caption[]{\label{fig1} {\em Schematic diagram for high energy diffractive
$J/\psi$ photoproduction.  The factorized form follows since, in
the proton rest frame, the scattering of the $c\overline{c}$
system occurs over a much shorter timescale than the $\gamma
\rightarrow c\overline{c}$ fluctuation or the $J/\psi$ formation
times.}} 
\end{center}
\end{figure}
To lowest-order the $\gamma^* p \rightarrow J/\psi p$ amplitude
can be factorized into the product of the $\gamma \rightarrow
c\overline{c}$ transition, followed by the scattering of the
$c\overline{c}$ system on the proton via (colourless) two-gluon
exchange, and finally the formation of the $J/\psi$ from the
outgoing $c\overline{c}$ pair.  The sequence is sketched in Fig.\ 1.  
The crucial observation is that at high energy the scattering
on the proton occurs over a much shorter timescale than the
$\gamma \rightarrow c\overline{c}$ fluctuation or the $J/\psi$
formation times.  Moreover the two-gluon exchange amplitude can
be shown to be directly proportional to the gluon density $g (x,
\overline{Q}^2)$ with
\begin{equation}
x \; = \; (M_\psi^2 + Q^2)/W^2, \;\;\;\;\; \overline{Q}^2 \; = \;
\textstyle{\frac{1}{4}} \: (M_\psi^2 + Q^2).
\label{a0}
\end{equation}
In view of the importance of this connection, the corrections to
the leading-order formula \cite{r,bfgms} have been studied
\cite{rrml,fks}.  It turns out that the major ambiguity is
associated with the
$J/\psi$ wave function.  In particular there are sizeable
normalization uncertainties which arise from allowing for the
relativistic motion of the $c$ and $\overline{c}$ quarks within the
$J/\psi$ meson.  Even though the normalization is not precisely 
determined, it is advocated \cite{rrml} that the \lq\lq shape" 
(or $W$ dependence) of the cross section for $J/\psi$ diffractive 
production at high energies can serve as a valuable probe of the small 
$x$ behaviour of the gluon.

From a theoretical point of view the study of diffractive open
charm production has some advantages as compared to
$J/\psi$ production.  It avoids the ambiguities associated
with the $J/\psi$ wave function and yet retains the {\it
quadratic} sensitivity to the gluon distribution.  Moreover, in
contrast to $J/\psi$, for open charm we can study the QCD
behaviour as a function of $M$, the invariant mass of the
$c\overline{c}$ system.  In principle due to the heavy quark mass, 
perturbative QCD can predict both the cross sections $\sigma^{L,T}$ for 
diffractive $c\overline{c}$ production from longitudinally and transversely 
polarised photons.  Indeed we find this to be the case.  We compute both 
the $Q^2$ and $M^2$ dependence of $\sigma_T$ and $\sigma_L$ by integrating 
over the transverse momenta $(k_T)$ of the produced $c$ and $\overline{c}$ 
quarks, and over the transverse momenta $(\ell_T)$ of the exchanged 
gluons.  Another feature is that the relevant scale at which the gluon is 
sampled in the diffractive production of open charm is
\begin{equation}
(m_c^2 \: + \: k_T^2) \; \left (1 \: + \: \frac{Q^2}{M^2} \right
),
\label{b0}
\end{equation}
which grows with $Q^2$.  Besides (lowest-order) $c\overline{c}$ production 
we estimate higher-order QCD corrections arising from real and virtual gluon 
emissions.  Recall that in the Drell-Yan process the ${\cal O} (\alpha_S)$ 
contributions contain $\pi^2$ factors and that on resummation of these terms 
we obtain a significant enhancement of the cross section.  In section 3 we 
will show that the ${\cal O} (\alpha_S)$ correction to 
diffractive $c\overline{c}$ production also has a $\pi^2$ enhancement. \\

\bigskip
\noindent {\large \bf 2.  The basic formulae for diffractive open
charm production}

Here we study the diffractive production of a $c\overline{c}$
pair of invariant mass $M$ from a photon of virtuality $Q^2$ at
high $\gamma p$ c.m.\ energy $\sqrt{s}$.  The lowest order
diagram for the process $\gamma^* p \rightarrow c\overline{c}p$
is shown in Fig.\ 2.  The Bjorken $x$ variable is given
by
\begin{equation}
x_B \; = \; Q^2/s.
\label{eq:a1}
\end{equation}
\begin{figure}[htb]
\begin{center}
\leavevmode
\epsfxsize=15.cm
\epsffile[100 340 540 550]{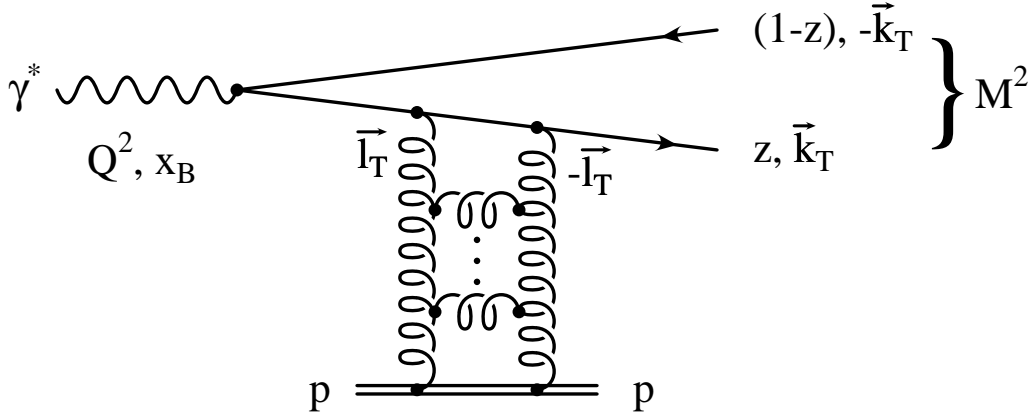}
\vskip -8mm
\caption[]{\label{fig2} {\em Diffractive open charm production in high energy
$\gamma^* p$ collisions, where $z$ is the fraction of the energy
of the photon that is carried by the charm quark.}} 
\end{center}
\end{figure}
As usual for a high energy diffractive process we introduce the
variables
\begin{equation}
x_{\funp} \; = \; \frac{Q^2 \: + \: M^2}{s},
\label{eq:a2}
\end{equation}
which is the longitudinal fraction of the proton energy carried
by the \lq\lq Pomeron" represented by the two-gluon exchange
ladder in Fig.\ 2, and
\begin{equation}
\beta \; = \; \frac{Q^2}{Q^2 \: + \: M^2} \; = \;
\frac{x_B}{x_{\funp}}.
\label{eq:a3}
\end{equation}
The transverse momenta of the outgoing $c$ and $\overline{c}$
quarks are denoted by $\pm \mbox{\boldmath $k$}_T$, and those of the
exchanged gluons by $\pm \mbox{\boldmath $\ell$}_T$; we are
studying near-forward production.

It is convenient to use light-cone perturbation theory (see, for
example, ref.\ \cite{bl}) and to express the particle
four momenta in the form
\begin{equation}
k_\mu \; = \; (k_+, k_-, \mbox{\boldmath $k$}_T)
\label{eq:a4}
\end{equation}
where $k_\pm = k_0 \pm k_3$.  In this approach all the particles
are on mass-shell, $k^2 = k_+ k_- - \mbox{\boldmath $k$}_T^2 =
m^2$, and $k_+$ and $\mbox{\boldmath $k$}_T$ are conserved at
each vertex.  We choose to work in a frame in which the target
proton is essentially at rest and where the other particles are
fast with four momenta
\begin{eqnarray}
\label{eq:a5}
q_\mu & = & (q_+, \; -Q^2/q_+, \; {\bf 0}), \nonumber
\\
k_\mu & = & (k_+, \; m_T^2/k_+, \; \mbox{\boldmath $k$}_T), \\
\ell_\mu & = & (\ell_+, \; \mbox{\boldmath $\ell$}_T^2/\ell_+, \;
\mbox{\boldmath $\ell$}_T), \nonumber
\end{eqnarray}
with 
$$
m_T^2 \equiv m^2 + \mbox{\boldmath $k$}_T^2,
$$
where $m$ is the mass of the charm quark.

\bigskip
\noindent {\bf 2.1.  Factorization of the diffractive $\gamma^*
\rightarrow c\overline{c}$ cross section}

The differential cross section for the diffractive production of
a $c\overline{c}$ pair of invariant mass $M$ is
\begin{equation}
\left . \frac{d^2 \sigma}{dM^2 dt} \right |_{t = 0} \; = \;
\sum_{\lambda, \lambda^\prime} \: \int \: \frac{d^2 k_T dz}{16
\pi^3} \; \delta \left ( M^2 \: - \: \frac{m_T^2}{z (1 - z)}
\right ) \; \frac{1}{16 \pi s^2} \: \left | {\cal M}_{\lambda
\lambda^\prime} \right |^2
\label{eq:a6}
\end{equation}
where ${\cal M}_{\lambda \lambda^\prime}$ is the amplitude for
the production of a $c$ and $\overline{c}$ (with helicities
$\lambda$ and $\lambda^\prime$) from the virtual photon.  The
$\delta$-function arises because the mass $M$ system is formed by
$c$ and $\overline{c}$ quarks with $k_+$ components $zq_+$ and
$(1 - z) q_+$.  We may rewrite (\ref{eq:a6}) in the form
\begin{equation}
x_{\funp} \: \left . \frac{d^2 \sigma}{dx_{\funp} dt} \right |_0
\;
= \;
\sum_{\lambda, \lambda^\prime} \: \frac{M^2 + Q^2}{M^2} \: \int
\: \frac{d^2 k_T dz}{16 \pi^3} \: \frac{m_T^2}{M^2} \; \delta
\left ( z (1 - z) \: - \: \frac{m_T^2}{M^2} \right ) \:
\frac{1}{16 \pi s^2} \; \left |{\cal M}_{\lambda \lambda^\prime}
\right |^2.
\label{eq:a7}
\end{equation}
As mentioned in the introduction, the high-energy diffractive
$c\overline{c}$ production amplitude ${\cal M}_{\lambda
\lambda^\prime}$ can be factorized into the light-cone wave function 
$\psi_{\lambda \lambda^\prime}$ of the 
$c\overline{c}$ pair in the virtual photon and the
(helicity-conserving) amplitude $T_{\lambda \lambda^\prime}$ for
the scattering of the $c\overline{c}$ pair on the target proton. 
In analogy to ref.\ \cite{bfgms} we have
\begin{equation}
{\cal M}_{\lambda \lambda^\prime} (k_T, z) \; = \; \sqrt{N_C} \:
\int \: d^2 k_T^\prime \: \int_0^1 \: dz^\prime
\: \psi_{\lambda \lambda^\prime} (k_T^\prime, z^\prime) \:
T_{\lambda \lambda^\prime} (k_T^\prime, z^\prime; k_T, z),
\label{eq:a8}
\end{equation}
where $\sqrt{N_C}$ occurs in the amplitude, since the cross section is the 
sum over the number of colours $N_C = 3$ of the charm quark. 

The factorization of ${\cal M}$ follows since the lifetime
$\tau_\gamma$ of the $c\overline{c}$ fluctuation of the virtual
photon is much longer than the time of interaction with the
gluons $\tau_i$.  It is informative to recall the argument of why
this is so.  According to the uncertainty principle the fluctuation
time
\begin{equation}
\tau_\gamma \: \sim \: \frac{1}{\Delta E} \; = \; \left |
\frac{1}{q_- - k_{1-} - k_{2-}} \right | \; = \; \frac{z (1 - z)
q_+}{\overline{Q}^2 + k_T^2}
\label{eq:b8}
\end{equation}
where $k_1$ and $k_2$ are the four momenta of the quarks of mass
$m$, and
\begin{equation}
\overline{Q}^2 \; \equiv \; z (1 - z) Q^2 \: + \: m^2.
\label{eq:c8}
\end{equation}
An estimate of the interaction time can be obtained from the
typical time for the emission of a gluon with momentum $\ell$,
from the quark $k_1$, say.  Then
\begin{equation}
\tau_i \: \sim \: \left | \frac{1}{k_{1-} - k_{1-}^\prime -
\ell_-} \right | \; = \; \left | \frac{q_+}{m_T^2/z -
m_T^2/z^\prime - \ell_T^2/\alpha} \right |
\label{eq:d8}
\end{equation}
where $\alpha = \ell_+/q_+$ and $z^\prime = z - \alpha$.  In the
leading $\log 1/x$ approximation we have $\alpha \ll z$ and hence
\begin{equation}
\tau_i \; \approx \; \frac{\alpha q_+}{\ell_T^2} \; \ll \;
\tau_\gamma.
\label{eq:e8}
\end{equation}
At high $Q^2$ the argument becomes particularly transparent. 
Then from (\ref{eq:b8}) we have
\begin{equation}
\tau_\gamma \: \sim \: \frac{q_+}{Q^2} \; = \; \frac{1}{m_p x_B},
\label{eq:f8}
\end{equation}
whereas (\ref{eq:e8}) gives
\begin{equation}
\tau_i \: \sim \: \frac{\ell_+}{\ell_T^2} \; = \; \frac{1}{m_p
x_\ell}
\label{eq:g8}
\end{equation}
where $x_\ell$ is the Bjorken variable for the gluon-proton
interaction and $m_p$ is the mass of the proton.  We are
concerned with the kinematic region $x_B \ll x_\ell$ where the
leading $\log (x_\ell/x_B)$ approximation is appropriate, and so
we have $\tau_\gamma \gg \tau_i$.

\begin{figure}[htb]
\begin{center}
\leavevmode
\epsfxsize=15.cm
\epsffile[20 220 570 590]{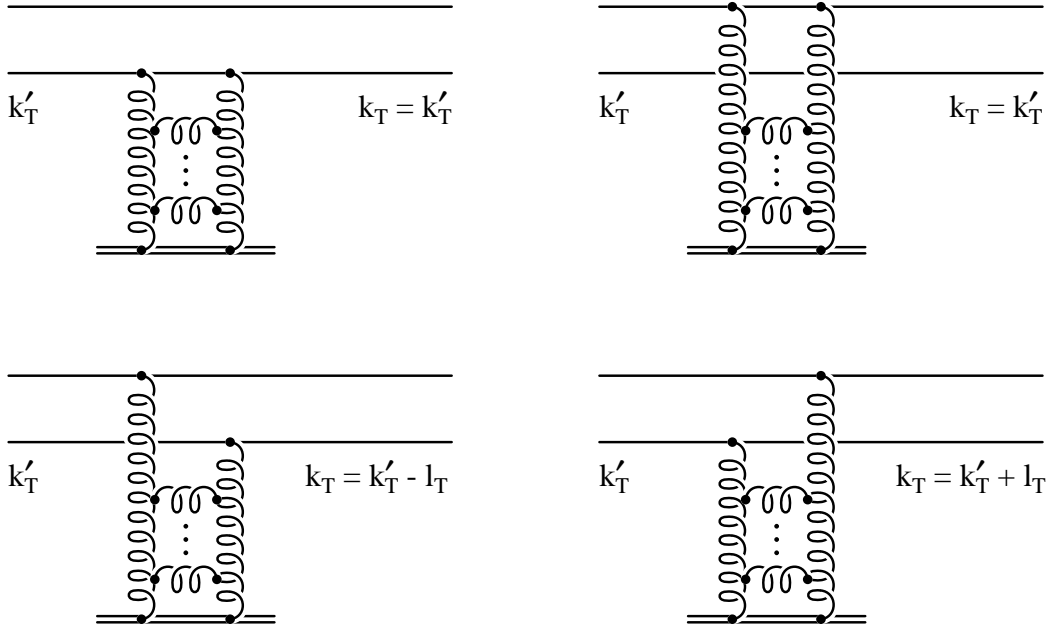}
\caption[]{\label{fig3} {\em Graphs which contribute to the amplitude
$T_{\lambda^\prime \lambda} (k_T^\prime, z^\prime; k_T, z)$ for
the scattering of the $c\overline{c}$ pair off the target proton.
At high energy $z = z^\prime$ and the quark helicities ($\lambda$
and $\lambda^\prime$) are conserved.}} 
\end{center}
\end{figure}
During the short interaction time $\tau_i$ the exchanged gluons
change only the quark (and/or antiquark) transverse momenta, and
leave their energy fractions and
helicities unchanged.  Thus $T_{\lambda \lambda^\prime}$ can be
simply computed as the sum of the light-cone perturbation theory
graphs shown in Fig.\ 3.  We have \cite{bfgms}
\begin{equation}
T_{\lambda \lambda^\prime} (k_T^\prime, k_T) \; = \;
i \: \frac{4\pi s}{2N_C} \: \int \: 
\frac{d^2 \ell_T}{\ell_T^4}
\; \biggl [ 2 \delta (\mbox{\boldmath $k$}_T^\prime -
\mbox{\boldmath $k$}_T) \: - \: \delta (\mbox{\boldmath
$k$}_T^\prime - \mbox{\boldmath $k$}_T - \mbox{\boldmath
$\ell$}_T) \: - \: \delta (\mbox{\boldmath $k$}_T^\prime - \mbox{\boldmath
$k$}_T + \mbox{\boldmath $\ell$}_T) \biggr ] \; \alpha_S (\ell_T^2) \: 
f (x_\funp, \ell_T^2),
\label{eq:a9}
\end{equation}
where in the leading $\ln (1/x)$ approximation, $z = z^\prime$ and $T_{\lambda 
\lambda^\prime}$ becomes independent of $z$ and $z^\prime$.  The factor 
$1/\ell_T^4$ arises from the propagators of the two exchanged gluons. 
The factor $1/2N_C$ arises from the colour coupling of the gluon to 
the quark and the $4\pi$ from the usual coupling 
relation $g_s^2 = 4\pi \alpha_S$.  The factor $s$ 
arises since our amplitude $T$ is defined so that the optical theorem reads 
\lq\lq Im $T = s\sigma$".  The distribution $f (x_\funp, \ell_T^2)$ is the 
probability of finding a $t$-channel 
gluon with transverse momentum $\ell_T$.  That is $f$ is the gluon density 
unintegrated over $\ell_T^2$ or, to be precise, $\ln \ell_T^2$.  It 
satisfies the BFKL equation, which effectively resums the leading 
$\alpha_S \ln (1/x)$ contributions.  We have tacitly assumed that we are 
considering a forward \lq\lq elastic" scattering amplitude with $t = 0$.  
However, the minimum value of $|t|$ is
\begin{equation}
t_{min} \; = \; \left ( \frac{Q^2 + M^2}{W^2} \: m_p \right )^2 \; \approx \; 
x_\funp^2 m_p^2.
\label{eq:b17}
\end{equation}
The $t_{min}$ effects are expected to be small even up to $x_\funp$ of about 
0.1 \cite{rrml}.  To relate $f$ to the conventional gluon density, which 
satisfies GLAP evolution, we must integrate over $\ell_T^2$.  We have
\begin{equation}
x_\funp g (x_\funp, Q^2) \; = \; \int^{Q^2} \: \frac{d\ell_T^2}{\ell_T^2} \: 
f (x_\funp, \ell_T^2)
\label{eq:c17}
\end{equation}
and the inverse relation
\begin{equation}
f (x_\funp, \ell_T^2) \; = \; \ell_T^2 \: \frac{\partial (x_\funp g (x_\funp, 
\ell_T^2))}{\partial \ell_T^2}.
\label{eq:d17}
\end{equation}
On substituting (\ref{eq:a9}) into (\ref{eq:a8}) we have
\begin{equation}
{\cal M}_{\lambda \lambda^\prime} \; = \; i \: \frac{4\pi^2 s}{2 \sqrt{N_C}} 
\: \int \: \frac{d \ell_T^2}{\ell_T^4} \; \Delta
\psi_{\lambda \lambda^\prime} \: \alpha_S (\ell_T^2) \: f (x_\funp, \ell_T^2),
\label{eq:a10}
\end{equation}
where $\Delta \psi$ is the combination
\begin{equation}
\Delta \psi (\mbox{\boldmath $k$}_T, \mbox{\boldmath $\ell$}_T,
z) \; = \; 2 \psi
(\mbox{\boldmath $k$}_T, z) \: - \: \psi (\mbox{\boldmath $k$}_T
- \mbox{\boldmath $\ell$}_T, z) \: - \:
\psi (\mbox{\boldmath $k$}_T + \mbox{\boldmath $\ell$}_T, z).
\label{eq:a11}
\end{equation}

To evaluate $\Delta \psi$ of (\ref{eq:a11}) we need the photon
wave function \cite{bfgms,m,nz}.  We use the convention of ref.\
\cite{bl} and express it in the form
\begin{equation}
\psi_{\lambda \lambda^\prime} (k_T, z) \; = \; - ee_c \: \sqrt{z
(1 - z)} \; \frac{\overline{u}_\lambda (k_T, z) \: \gamma \cdot 
\epsilon \: v_{\lambda^\prime} (- k_T, 1 - z)}{\overline{Q}^2 +
k_T^2}
\label{eq:a12}
\end{equation}
where $ee_c = \frac{2}{3}e$ is the charge of the charm quark, 
$\epsilon$ is the polarisation vector of the photon, and
$\overline{Q}^2$ is given by (\ref{eq:c8}).

\bigskip
\noindent {\bf 2.2.  The helicity amplitudes for diffractive $c\overline{c}$ 
production}

In this subsection we explicitly evaluate the helicity amplitudes 
${\cal M}_{\lambda \lambda^\prime}^{\lambda (\gamma)}$ which describe the 
diffractive production of a $c\overline{c}$ pair with helicities $\lambda, 
\lambda^\prime$ from a photon of helicity $\lambda (\gamma)$.  First we obtain 
the amplitudes for production from a longitudinally polarised photon and then 
from a transversely polarised photon.

The four momentum of the photon has the form $q_\mu = (q_+, \:
-Q^2/q_+, \: {\bf 0})$, see (\ref{eq:a5}), and thus a
longitudinally polarised photon is described by the polarisation
vector
\begin{equation}
\epsilon_L \; = \; (q_+/Q, \: Q/q_+, {\bf 0}).
\label{eq:a14}
\end{equation}
In order to evaluate (\ref{eq:a12}) with $\epsilon = \epsilon_L$
we first note that
$$
q^\mu \overline{u}_\lambda \gamma_\mu v_{\lambda^\prime} \; = \;
\frac{1}{2} \: (q_+ \overline{u}_\lambda \gamma_-
v_{\lambda^\prime} \; + \; q_- \overline{u}_\lambda \gamma_+
v_{\lambda^\prime}) \; = \; 0
$$
so that we have
\begin{equation}
\overline{u} \gamma_- v \; = \; \frac{Q^2}{q_+^2} \: \overline{u}
\gamma_+ v.
\label{eq:a15}
\end{equation}
We may use this identity to evaluate the matrix element which
occurs in (\ref{eq:a12})
\begin{eqnarray}
\overline{u}_\lambda \: \gamma . \epsilon \: v_{\lambda^\prime} &
= & \frac{1}{2} \: \left ( \frac{q_+}{Q} \: \overline{u}_\lambda
\gamma_- v_{\lambda^\prime} \: + \:  \frac{Q}{q_+} \:
\overline{u}_\lambda \gamma_+ v_{\lambda^\prime} \right ) \; = \;
\frac{Q}{q_+} \: \overline{u}_\lambda \gamma_+ v_{\lambda^\prime}
\nonumber \\
& & \nonumber \\
& = & 2 \: Q \: \sqrt{z (1 - z)} \: \delta_{\lambda,
-\lambda^\prime}
\label{eq:a16}
\end{eqnarray}
see, for example, ref.\ \cite{bl}.  Using (\ref{eq:a12}) and
(\ref{eq:a16}) we can evaluate $\Delta \psi$ for longitudinally
polarised photons.  We obtain
\begin{equation}
\Delta \psi_{\lambda \lambda^\prime}^L \; = \; -2
\delta_{\lambda, - \lambda^\prime} \: ee_c \: Q \: z (1-z)
\; \left [ \frac{2}{\overline{Q}^2 + k_T^2} \: - \:
\frac{1}{\overline{Q}^2 + (\mbox{\boldmath $k$}_T -
\mbox{\boldmath $\ell$}_T)^2} \: - \: \frac{1}{\overline{Q}^2 +
(\mbox{\boldmath $k$}_T + \mbox{\boldmath $\ell$}_T)^2} \right ].
\label{eq:a17}
\end{equation}
We substitute $\Delta \psi^L$ in (\ref{eq:a10}) and carry out the 
angular integration.  We find the amplitudes for the production of $c$ and 
$\overline{c}$ quarks with helicities $\lambda, \lambda^\prime$ from a 
longitudinally polarised photon to be 
\begin{equation}
{\cal M}_{\lambda \lambda^\prime}^L \; = \; - 4 \delta_{\lambda, 
- \lambda^\prime} \: ee_c \: Q \: z (1 - z) \; \left ( i \: 
\frac{2\pi^2 s}{\sqrt{N_C}} \right ) \: {\cal I}_1 (k_T^2)
\label{eq:b24}
\end{equation}
where the integral over the transverse momenta of the exchanged gluons is
\begin{equation}
{\cal I}_1 (k_T^2) \; = \; \int \: \frac{d\ell_T^2}{\ell_T^4} \: \alpha_S 
(\ell_T^2) \: f (x_\funp, \ell_T^2) \: \left (\frac{1}{\overline{Q}^2 + 
k_T^2} \: - \: \frac{1}{Q_{k\ell}^2} \right ).
\label{eq:c24}
\end{equation}
Here we have introduced
\begin{equation}
Q_{k\ell}^2 \; = \; \sqrt{(\overline{Q}^2 + k_T^2 + \ell_T^2)^2 \: - \: 4k_T^2 
\ell_T^2}.
\label{eq:d24}
\end{equation}

In order to calculate the helicity amplitudes for diffractive $c\overline{c}$ 
production from a transversely polarised incoming photon it is convenient to 
evaluate \cite{m,bl} the matrix
element $\overline{u} \mbox{\boldmath $\gamma$}_T \cdot
\mbox{\boldmath $\epsilon$}_T v$ using as a basis the circular
polarisation vectors of the photon
\begin{equation}
\epsilon_\pm \; = \; \frac{1}{\sqrt{2}} \: (0, 0, 1, \pm i).
\label{eq:a19}
\end{equation}
Then we obtain
\begin{equation}
\overline{u}_\lambda (\mbox{\boldmath $\gamma$}_T .
\mbox{\boldmath $\epsilon$}_\pm) v_{\lambda^\prime} \; = \;
\frac{1}{\sqrt{z (1 - z)}} \: \biggl ( \delta_{\lambda, -
\lambda^\prime} \left \{ (1 - 2z) \lambda \: \overline{+} 1
\right \} \: \mbox{\boldmath $\epsilon$}_\pm \cdot
\mbox{\boldmath $k$}_T \: + \: \delta_{\lambda
\lambda^\prime} \: m \lambda
\biggr ),
\label{eq:a20}
\end{equation}
where here quark helicities of $\pm \frac{1}{2}$ are represented
by $\lambda = \pm 1$.  In this polarisation basis we have
\begin{eqnarray}
\Delta \psi_{\lambda \lambda^\prime}^\pm & = & - ee_c \:
\delta_{\lambda, - \lambda^\prime} \left \{ (1 - 2z) \lambda
\overline{+} 1 \right \} \; \left [ \frac{2 \mbox{\boldmath
$\epsilon$}_\pm \cdot \mbox{\boldmath $k$}_T}{\overline{Q}^2 +
k_T^2} \: - \: \frac{\mbox{\boldmath $\epsilon$}_\pm \cdot 
(\mbox{\boldmath $k$}_T - \mbox{\boldmath
$\ell$}_T)}{\overline{Q}^2 + (\mbox{\boldmath $k$}_T -
\mbox{\boldmath $\ell$}_T)^2} \: - \:
\frac{\mbox{\boldmath $\epsilon$}_\pm \cdot (\mbox{\boldmath
$k$}_T + \mbox{\boldmath $\ell$}_T)}{\overline{Q}^2 +
(\mbox{\boldmath $k$}_T + \mbox{\boldmath $\ell$}_T)^2} \right ]
\nonumber \\
& & \nonumber \\
& & - ee_c \: \delta_{\lambda \lambda^\prime} \: m
\lambda \; \left [ \frac{2}{\overline{Q}^2 + k_T^2} \: - \:
\frac{1}{\overline{Q}^2 + (\mbox{\boldmath $k$}_T -
\mbox{\boldmath $\ell$}_T)^2} \: - \: \frac{1}{\overline{Q}^2 +
(\mbox{\boldmath $k$}_T + \mbox{\boldmath $\ell$}_T)^2} \right ].
\label{eq:a21}
\end{eqnarray}
As before we substitute $\Delta \psi^\pm$ into (\ref{eq:a10}) and carry 
out the angular integration.  We obtain the helicity amplitudes
\begin{equation}
{\cal M}_{\lambda \lambda^\prime}^\pm \; = \; -2 ee_c \: \left ( i \: 
\frac{2\pi^2 s}{\sqrt{N_C}} \right ) \: \biggl (\delta_{\lambda, 
- \lambda^\prime} \left \{ (1 - 2z) \lambda \overline{+} 
1 \right \} \: \mbox{\boldmath $\varepsilon$}_\pm . \mbox{\boldmath $k$}_T \: 
{\cal I}_2 (k_T^2) \: + \: \delta_{\lambda \lambda^\prime} m\lambda \: 
{\cal I}_1 (k_T^2) \biggr )
\label{eq:b28}
\end{equation}
where the integral ${\cal I}_1$ is given by (\ref{eq:c24}) and
\begin{equation}
{\cal I}_2 (k_T^2) \; = \; \int \: \frac{d\ell_T^2}{\ell_T^4} \: \alpha_S 
(\ell_T^2) \: f (x_\funp, \ell_T^2) \; 
\left ( \frac{1}{\overline{Q}^2 + k_T^2} 
\: - \: \frac{1}{2k_T^2} \: + \: \frac{\overline{Q}^2 - k_T^2 + \ell_T^2}
{2k_T^2 Q_{k\ell}^2} \right )
\label{eq:c28}
\end{equation}
with $\overline{Q}^2$ and $Q_{k\ell}^2$ defined as in (\ref{eq:c8}) and 
(\ref{eq:d24}) respectively.

Asymptotically, when $\overline{Q}^2 + k_T^2 \gg \ell_T^2$, the $\ell_T^2$ 
integration gives a logarithmic contribution of the form $\ln [(\overline{Q}^2 
+ k_T^2)/\mu^2]$, or rather like $1/\gamma$ if the anomalous 
dimension $\gamma$ of the gluon distribution is taken into account.  At 
realistic energies non-logarithmic contributions are appreciable and the 
integrations ${\cal I}_{1,2} (k_T^2)$ have to be performed explicitly.  The 
important domain of integration is $\mu^2 \lapproxeq \ell_T^2 \lapproxeq 
\overline{Q}^2 + k_T^2$.

\bigskip
\noindent {\bf 2.3.  The diffractive $\gamma_{T,L} \rightarrow
c\overline{c}$ cross sections}

To evaluate the cross sections for open charm production we need to 
sum $|{\cal M}|^2$ over the quark helicities $\lambda$ and $\lambda^\prime$.  
For production from longitudinally polarised photons we have 
from (\ref{eq:b24})
\begin{equation}
\sum_{\lambda, \lambda^\prime} \left | {\cal M}_{\lambda\lambda^\prime}^L 
\right |^2 \; = \; \frac{4\pi^4 s^2}{3} \: 32 \: e^2 e_c^2 \: z^2 (1 - z)^2 \: 
Q^2 \: | {\cal I}_1 |^2,
\label{eq:d28}
\end{equation}
while for transversely polarised photons we must average over the two 
transverse polarisation states $\lambda (\gamma) = \pm$. 
From (\ref{eq:b28}) we have
\begin{equation}
\frac{1}{2} \: \sum_{\lambda (\gamma) = \pm} \: \sum_{\lambda, 
\lambda^\prime} \: 
\left | {\cal M}_{\lambda, \lambda^\prime}^{\lambda (\gamma)} 
\right |^2 \; = \; 
\frac{4\pi^4 s^2}{3} \: 8 e^2 e_c^2 \: \biggl ( k_T^2 \{z^2 + (1 - z)^2 \} \: 
|{\cal I}_2 |^2 \: + \: m^2 \: |{\cal I}_1 |^2 \biggr ).
\label{eq:e28}
\end{equation}

We are now ready to calculate the cross sections from (\ref{eq:a7}).  On 
carrying out the $z$ integration in (\ref{eq:a7}) we find that
the $\delta$ function gives a Jacobian factor $2 (1-4
m_T^2/M^2)^{- \frac{1}{2}}$, and moreover implies that
\begin{eqnarray}
z (1-z) & = & \frac{m_T^2}{M^2}, \nonumber \\
& & \nonumber \\
z^2 \: + \: (1-z)^2 & = & 1 \: - \: \frac{2m_T^2}{M^2} \; = \; 1
\: - \: \frac{2 \beta}{1 - \beta} \: \frac{m_T^2}{Q^2} \; \equiv
\; 1 \: - \: \frac{2 \beta K^2}{Q^2}, \nonumber 
\end{eqnarray}
where we have defined
\begin{equation}
K^2 \; \equiv \; \frac{m^2 + k_T^2}{1 - \beta}.
\label{eq:a26}
\end{equation}
The variable $K^2$ turns out to be the scale probed by the
process, since after the $z$ integration
\begin{equation}
\overline{Q}^2 \: + \: k_T^2 \; = \; Q^2 \: \frac{m_T^2}{M^2} \:
+ \: m_T^2 \; = \; \frac{m_T^2}{1 - \beta} \; \equiv \; K^2.
\label{eq:a27}
\end{equation}
Collecting the factors together and inserting into (\ref{eq:a7}) we find 
that the cross section for diffractive open charm production from
longitudinally polarised photons is
\begin{equation}
x_{\funp} \: \left . \frac{d^2 \sigma^L}{dx_{\funp} dt} \right
|_0 \; = \;
\frac{4 \pi^2 e_c^2 \alpha}{3} \; \frac{Q^2}{1 - \beta} \:
\int_{0}^{\frac{1}{4} M^2 - m^2} \frac{dk_T^2}{\sqrt{1 - 4
m_T^2/M^2}} \: \left ( \frac{m_T^2}{M^2} \right )^3 \: | {\cal I}_1 |^2,
\label{eq:a28}
\end{equation}
and from transversely polarised photons is
\begin{equation}
x_{\funp} \: \left . \frac{d^2 \sigma^T}{dx_{\funp} dt}
\right |_0 \; =
\; \frac{\pi^2 e_c^2 \alpha}{3 (1 - \beta)} \:
\int_{0}^{\frac{1}{4} M^2 - m^2} \: \frac{dk_T^2}{\sqrt{1 - 4
m_T^2/M^2}} \: \frac{m_T^2}{M^2} \: \left ( k_T^2 \: \left ( 1 - 
\frac{2\beta K^2}{Q^2} \right ) \: |{\cal I}_2 |^2 \: + \: m^2 
|{\cal I}_1 |^2 \right )
\label{eq:a29}
\end{equation}
where recall that $m_T^2 \equiv m^2 + k_T^2$ and $\beta =
Q^2/(Q^2 + M^2)$.  The masses $m$ and $M$ are those of the $c$ quark
and the $c\overline{c}$ system respectively, and $\alpha$ is the
electromagnetic coupling. 

\bigskip
\noindent {\bf 2.4.  Leading logarithmic approximation}

In practice the ${\cal I}_{1,2}$ integrations 
over $\ell_T^2$ of (\ref{eq:c24}) 
and (\ref{eq:c28}) have to be performed explicitly, with the main contribution 
arising from the domain $\ell_T^2 \lapproxeq \overline{Q}^2 + k_T^2$. 
However, before we do this, it is informative to derive analytical 
expressions for the integrals assuming that 
the main contribution comes from the region $\ell_T^2 \ll \overline{Q}^2 + 
k_T^2$.  That is we expand the terms in brackets in (\ref{eq:c24}) and 
(\ref{eq:c28}) in the form $c_1 \ell_T^2 + c_2 \ell_T^4 + \ldots$ and retain 
only the $c_1$ term.  In this approximation ${\cal I}_{1,2}$ are given by
\begin{eqnarray}
\label{eq:b29}
{\cal I}_1^{LLA} (k_T^2) & = & \frac{\overline{Q}^2 - k_T^2}{(\overline{Q}^2 
+ k_T^2)^3} \: \alpha_S (\overline{Q}^2 + k_T^2) \: x_\funp g (x_\funp, 
\overline{Q}^2 + k_T^2) ,\\
& & \nonumber \\
\label{eq:c29}
{\cal I}_2^{LLA} (k_T^2) & = & \frac{2 \overline{Q}^2}{(\overline{Q}^2 
+ k_T^2)^3} \: \alpha_S (\overline{Q}^2 + k_T^2) \: x_\funp g (x_\funp, 
\overline{Q}^2 + k_T^2)
\end{eqnarray}
where we have used (\ref{eq:c17}) to express the answer in terms of the 
conventional gluon distribution.

If we use these results to evaluate $|{\cal M}|^2$ and substitute them into 
(\ref{eq:a7}) then we find, in this approximation, that the cross sections are 
given by
\begin{eqnarray}
\label{eq:d29}
& & x_{\funp} \: \left . \frac{d^2 \sigma^L}{dx_{\funp} dt} \right
|_0 \; = \nonumber \\
& & \;\;\;\;\; \frac{4 \pi^2 e_c^2 \alpha}{3 Q^4} \; \beta^3 \:
\int_{0}^{\frac{1}{4} M^2 - m^2} \frac{dk_T^2}{\sqrt{1 - 4
m_T^2/M^2}} \: \frac{[m_T^2 - 2 (1 - \beta) k_T^2 ]^2}{(m_T^2)^3}
\biggl ( \alpha_S (K^2) \: x_{\funp} g (x_{\funp}, K^2) \biggr
)^2, \nonumber \\
& & \\
& & \nonumber \\
& & x_{\funp} \: \left . \frac{d^2 \sigma^T}{dx_{\funp} dt}
\right |_0 \; =
\; \frac{4 \pi^2 e_c^2 \alpha}{3 Q^2} \: \beta (1 - \beta)^2 \:
\int_{0}^{\frac{1}{4} M^2 - m^2} \: \frac{dk_T^2}{\sqrt{1 - 4
m_T^2/M^2}} \nonumber \\
& & \nonumber \\
\label{eq:e29}
& & \left \{ \frac{k_T^2 (m^2 + \beta k_T^2)^2 \: (1 - 2 \beta
K^2/Q^2)}{(m_T^2)^5} \: + \: \frac{m^2}{4} \: \frac{[m_T^2 - 2 (1
- \beta) k_T^2]^2}{(m_T^2)^5} \right \} \; \biggl ( \alpha_S
(K^2) \: x_{\funp} g (x_{\funp}, K^2) \biggr )^2, \nonumber \\
& &
\end{eqnarray}
where the scale $K^2$ is given by (\ref{eq:a26}).

The scale $K^2$ for diffractive processes has been emphasized by Bartels 
et al.\ \cite{blw} and by Genovese et al.\ \cite{gnz}.  Ref.\ \cite{blw} 
concerns light quark-antiquark production and the \lq\lq hardness" of the 
scale $K^2$, and the validity of perturbative QCD is ensured by considering 
quark jets at large $k_T^2$.  Ref.\ \cite{gnz} discusses open charm 
production, based on earlier work starting from ref.\ \cite{nz2}.  It 
contains similar 
cross section formulae to (\ref{eq:a28}) and (\ref{eq:a29}), together with the 
logarithmic approximations of the form of (\ref{eq:d29}) and (\ref{eq:e29}).  
In particular the hardness of the scale $K^2 = (m^2 + k_T^2)/(1 - \beta)$ is 
emphasized, although the numerical predictions are made using the leading 
logarithmic approximation only, with the scale $K^2$ set equal 
to $m^2$.  Here, 
in section 4, we extend the numerical treatment to calculate explicitly the 
integrals ${\cal I}_{1,2}$ over the exchanged gluon transverse 
momenta $\ell_T$, 
and moreover we evaluate the higher order contributions (of section 3).  We 
shall see that the predicted values of the cross sections for diffractive 
$c\overline{c}$ production are considerably enhanced by both of these effects.

\bigskip
\noindent {\bf 2.5.  Connection with diffractive $J/\psi$
production}

It is instructive to see how these approximate cross section formulae given in
(\ref{eq:d29}) and (\ref{eq:e29}) compare with the result which
was derived for the exclusive production of heavy quark-antiquark
mesons [2-4].  A convenient way to make the comparison is to
consider the production of the mesonic states with $M^2 = M_V^2
\ll \tilde{Q}^2$ where
\begin{equation}
\tilde{Q}^2 \; \equiv \; Q^2 \: + \: M_V^2.
\label{eq:b39}
\end{equation}
The kinematic region for such a reaction corresponds to
\begin{equation}
\beta \; = \; 1 \: - \: \frac{M_V^2}{\tilde{Q}^2} \; \rightarrow
\; 1
\label{eq:c39}
\end{equation}
at large $\tilde{Q}^2/M_V^2$.  In this kinematic region we can
rewrite the cross section formulae (\ref{eq:d29}) and
(\ref{eq:e29}) in the form
\begin{eqnarray}
\label{eq:d39}
M^2 \left . \frac{d^2 \sigma^L}{dM^2 dt} \right |_0 & = & (1 -
\beta) \: x_{\funp} \: \left . \frac{d^2 \sigma^L}{dx_{\funp} dt}
\right |_0
\nonumber \\
& & \nonumber \\
& = & \frac{4 \pi^2 e_c^2 \alpha}{3} \; \frac{M_V^2
Q^2}{\tilde{Q}^8} \: \int_{0}^{\frac{1}{4} M^2 - m^2}
\frac{dk_T^2}{\sqrt{1 - 4 m_T^2/M^2}} \; \frac{1}{m_T^2} \;
\biggl ( \alpha_S (K^2) \: x_{\funp} \: g (x_{\funp}, K^2) \biggr
)^2
\nonumber \\
& & \nonumber \\
& \approx & \frac{4 \pi^2 e_c^2 \alpha}{3} \;
\frac{M_V^2 Q^2}{(Q^2 + M_V^2)^4} \; \biggl (\alpha_S
(\overline{Q}^2) \: x_{\funp} \: g (x_{\funp}, \overline{Q}^2)
\biggr )^2 \: {\cal I}_L, \\
& & \nonumber \\
\label{eq:e39}
M^2 \left . \frac{d^2 \sigma^T}{dM^2 dt} \right |_0 & = &
\frac{4 \pi^2 e_c^2 \alpha}{3} \;
\frac{M_V^4}{\tilde{Q}^8} \int_{0}^{\frac{1}{4} M^2 - m^2}
\frac{dk_T^2}{\sqrt{1 - 4 m_T^2/M^2}} \; \frac{M_V^2 (m^2 +
2k_T^2)}{4 (m_T^2)^3} \; \biggl (\alpha_S (K^2) \: x_{\funp} g
(x_{\funp},
K^2) \biggr )^2 \nonumber \\
& & \nonumber \\
& \approx & \frac{4 \pi^2 e_c^2 \alpha}{3} \; \frac{M_V^4}{(Q^2 +
M_V^2)^4} \; \biggl (\alpha_S (\overline{Q}^2) \: x_{\funp} \: g
(x_{\funp},
\overline{Q}^2) \biggr )^2 \: {\cal I}_T
\end{eqnarray}
where the scale $\overline{Q}^2$ is obtained by neglecting
$k_T^2$ in $K^2$, that is $\overline{Q}^2 \approx \frac{1}{4}
(Q^2 + M_V^2)$, and where ${\cal I}_{L,T}$ denote the integrals
over $k_T^2$.  For the exclusive diffractive production of a
meson of mass $M_V$ the integrals ${\cal I}_{L,T}$ should be
replaced by the overlap integrals of the
virtual photon wave function with light-cone wave function of
the $c\overline{c}$ pair in the meson [2-4].

Formulae (\ref{eq:d39}) and (\ref{eq:e39}) have precisely the
structure found for the exclusive diffractive production of a
meson of mass $M_V$ (see, for example, the $J/\psi$ diffractive
production cross section given in eq.\ (2) of ref.\ \cite{rrml}),
including the result
\begin{equation}
\sigma_L/\sigma_T \; \sim \; Q^2/M_V^2.
\label{eq:f39}
\end{equation}
However, to obtain reliable estimates of $\sigma_{L,T} (J/\psi)$
we need to evaluate the overlap integrals which replace ${\cal
I}_{L,T}$ in (\ref{eq:d39}) and (\ref{eq:e39}).  We return now to 
the subject of this paper, namely the diffractive production of open charm. \\

\bigskip
\noindent {\large \bf 3.  Higher order contributions to diffractive 
$c\overline{c}$ production}

So far we have considered only lowest-order diffractive $c\overline{c}$ 
production.  In this section we evaluate higher-order QCD corrections to the 
cross sections.  

\bigskip
\noindent {\bf 3.1.  Real gluon emission contributions}

We wish to calculate the contributions in which the incoming photon 
produces a $c\overline{c}g$ three jet configuration.  As before, the time 
during which the two gluon exchange interacts with the 
$c\overline{c}g$ system is much less than the lifetime of the $c\overline{c}g$ 
fluctuation of the photon.  So we can write the amplitude in factorized form 
and consider just the coupling of the exchanged gluons to the $c\overline{c}g$ 
state.

\begin{figure}[htb]
\begin{center}
\leavevmode
\epsfxsize=15.cm
\epsffile[80 400 550 590]{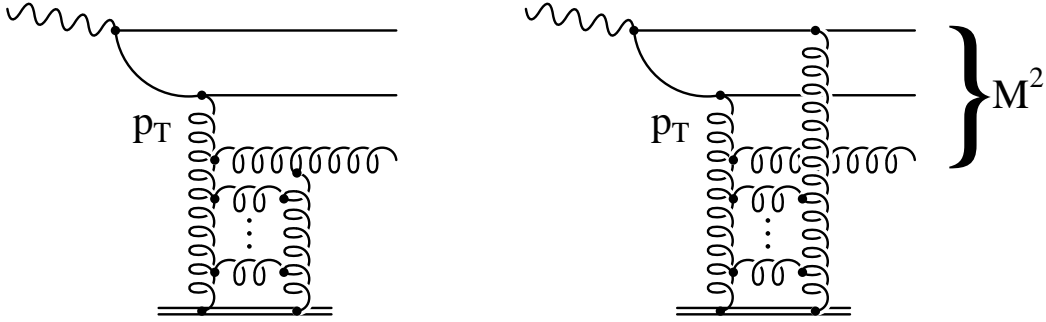}
\vskip -8mm
\caption[]{\label{fig4} {\em $c\overline{c}g$ contributions with strong 
ordering with the gluon having much smaller transverse momentum than the $c$ 
and $\overline{c}$.}} 
\end{center}
\end{figure}

We need to consider only diagrams with space-like evolution.  Those with 
time-like evolution do not change the cross section apart from the 
corrections discussed in the next subsection.  We divide the 
contributions according to the transverse momenta ordering of 
the $c, \overline{c}$ and $g$.  First we have contributions when the 
transverse momentum of the final gluon is much less than that of the
quarks.  The two 
exchanged gluons can couple to any of the three outgoing partons.  There are 
eighteen such diagrams, two of which are shown in Fig.\ 4.  These 
diagrams initiate 
GLAP evolution of a diffractive state starting from a gluon of transverse 
momentum $p_T$.  They give a $dM^2/M^2$ behaviour for $M^2 \gg Q^2, m^2$.  In 
contrast in the kinematic region where the outgoing gluon has 
transverse momenta 
greater than one of the quarks we have a $dM^2/M^4$ behaviour.  Fig.\ 5 shows 
such a contribution.  It corresponds to GLAP evolution of a diffractive 
state starting from an initial quark (with transverse 
momentum $p_T$).  Strictly 
speaking we should calculate the $c\overline{c}g$ contributions 
over all regions of phase space.  At present the required formulae 
only exist for strong ordering of the transverse momenta.  However, this 
should give a good estimate.  In fact 
the situation is better than it first appears.  For the low $M^2$ region 
lowest-order $c\overline{c}$ production is dominant.  For higher $M^2$ we 
can use the so-called Pomeron-gluon splitting function, $P_{gP}$, which 
is given by \cite{lw} 
\begin{equation}
P_{gP} (z) \; = \; \left ( \frac{8 N_C^2}{N_C^2 - 1} \right ) \:
\frac{1}{z} \: (1 + 2z)^2 \: (1 - z)^3,
\label{eq:a44}
\end{equation}
which is valid either for transverse-momentum ordering (with the outgoing 
gluon having the smallest transverse momentum) or in the BFKL limit of large 
$\ln (1/z)$.  The latter corresponds to large $M^2$ where the $c\overline{c}g$ 
configuration becomes important.

\begin{figure}[htb]
\begin{center}
\leavevmode
\epsfxsize=8.cm
\epsffile[200 360 440 600]{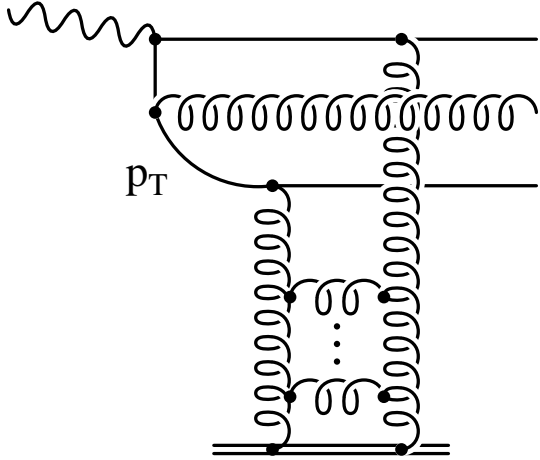}
\vskip -8mm
\caption[]{\label{fig5} {\em The $c\overline{c}g$ diagram driven by 
the quark distribution $(p_T)$.}} 
\end{center}
\end{figure}

Using the formulae of ref.\ \cite{lw} with the coefficient functions of 
refs.\ \cite{gr,w} we find
\begin{equation}
x_{\funp} \: \left . \frac{d^2 \sigma^{T,L}}{d x_{\funp} dt} \right |_0 \; =
\; \frac{4 \pi^2 e_c^2 \alpha}{Q^2} \: \frac{\alpha_S^3
(\mu^2)}{64 \pi} \; 2\beta \: \int_{z_m}^1 \: \frac{dz}{z} \; 
C^{T,L} \left ( \frac{\beta}{z}, \: \frac{m^2}{Q^2}
\right ) \: P_{gP} (z) \: \int_{Q_0^2}^{\mu^2} \:
\frac{dk_T^2}{k_T^4} \: \biggl ( x_{\funp} g (x_{\funp}, k_T^2)
\biggr )^2
\label{eq:a40}
\end{equation}
where the lower limit $z_m$ is determined by the kinematical
boundary
\begin{equation}
z_m \; = \; \beta \: \left ( 1 \: + \: \frac{4 m^2}{Q^2} \right
) \; = \; \frac{Q^2 + 4 m^2}{Q^2 + M^2},
\label{eq:b40}
\end{equation}
and the cut-off $Q_0^2$ is chosen to be 1 GeV$^2$.  The
leading-order coefficient functions are given by
\begin{eqnarray}
\label{eq:a41}
C^T (x, r) \: + \: C^L (x, r) & = & \biggl [ x^2 \: + \: (1 - x)^2 \:
+ \: 4 x (1 - 3x) r \: - \: 8x^2 r^2 \biggr ] \; \ln \: \frac{1 + v}{1 - v}
\nonumber \\
& & \nonumber \\
& & + \; v \: \biggl [ - 1 \: + \: 8x (1 - x) \: - \: 4x (1 - x)
r \biggr ], \\
& & \nonumber \\
\label{eq:a42}
C^L (x, r) & = & 4 v \: x (1 - x) \: - \: 8 x^2 r \: \ln \:
\frac{1 + v}{1 - v},
\end{eqnarray}
where $r = m^2/Q^2$ and $v$, the c.m.\ velocity of the $c$ quarks, is given by
\begin{equation}
v^2 \; = \; 1 \: - \: \frac{4 m^2 x}{Q^2 (1 - x)}.
\label{eq:a43}
\end{equation}
We take the mass factorization scale $\mu^2 = 4 m^2$, since
it was shown \cite{grs} that this is the most appropriate choice
for the perturbative stability of (\ref{eq:a40}).

From (\ref{eq:a41}) and (\ref{eq:a42}) we can gain insight into
how gluon emission contributes to the evolution equation.  For
large $Q^2/m^2$ we can rewrite (\ref{eq:a41}) and
(\ref{eq:a42}) in the form
\begin{eqnarray}
\label{eq:b50}
C^T (x, r) & = & [x^2 + (1 - x)^2 ] \: \ln \: \frac{Q^2}{m^2}
\nonumber \\
& & \nonumber \\
& & + \; \left \{ - 1 \: + \: 4x (1 - x) \: + \: \left [x^2 \: +
\: (1 - x)^2 \right ] \; \ln \: \left (\frac{1 - x}{x} \right )
\right \} , \\
& & \nonumber \\
\label{eq:c50}
C^L (x, r) & = & 4x (1 - x) .
\end{eqnarray}

\noindent Since $z_m \rightarrow \beta$ in the limit of large
$Q^2/m^2$,
we see that the $\log (Q^2/m^2)$ term in (\ref{eq:b50})
generates the usual GLAP evolution equation, with the appropriate
splitting function, for the diffractive dissociation structure
function (as given in (\ref{eq:a45}) below).

In principle it is straightforward to also evolve from the quark starting 
distribution and to determine its contribution at large $Q^2$.  However, it 
should be a small correction in the HERA regime.

\bigskip
\noindent {\bf 3.2.  Virtual contributions}

So far we have discussed higher order contributions arising from real gluon 
emission.  We should also study virtual loop corrections.  For the Drell-Yan 
process such contributions change the cross section by a factor of 2 or more 
--- the famous $K$ factor.  We must 
therefore investigate whether or not a similar enhancement occurs in the 
diffractive production of open charm.  Unfortunately at present there are no 
complete calculations of the ${\cal O} (\alpha_S)$ corrections for the 
diffractive process.  However, it is well known that a large (usually 
the dominant) part 
of the $K$ factor (say, in the Drell-Yan $q\overline{q} \rightarrow \gamma^*$ 
process \cite{dyk}) comes from the 
analytical continuation of the Sudakov form factor \cite{sud}, which leads to 
a contribution proportional to $(i \pi)^2$.  This contribution comes from the 
product of two imaginary parts --- that is the discontinuities shown 
by the dashed lines in Fig.\ 6a which each give a factor of $i \pi$.  It 
is closely related to the double logarithmic contribution. 

At first sight it appears that there will be no $\pi^2$ enhancements to the 
${\cal O} (\alpha_S)$ corrections to diffractive $c\overline{c}$ production, 
which from a simplified viewpoint looks like the process $\gamma^* \rightarrow 
c\overline{c}$.  If this simple view were true then to ${\cal O} (\alpha_S)$ 
the cross section would be $\sigma_0 (1 + \alpha_S/\pi)$, and there are 
manifestly no $\pi^2 \alpha_S/\pi$ terms apart of course from the 
usual threshold corrections.  However, as we shall see, when we allow 
for the two-gluon exchange interaction we find that there are new 
diagrams which lead to a $\pi^2$ enhancement.

\begin{figure}[htb]
\begin{center}
\vskip -6mm
\leavevmode
\epsfxsize=17.cm
\epsffile[70 360 570 560]{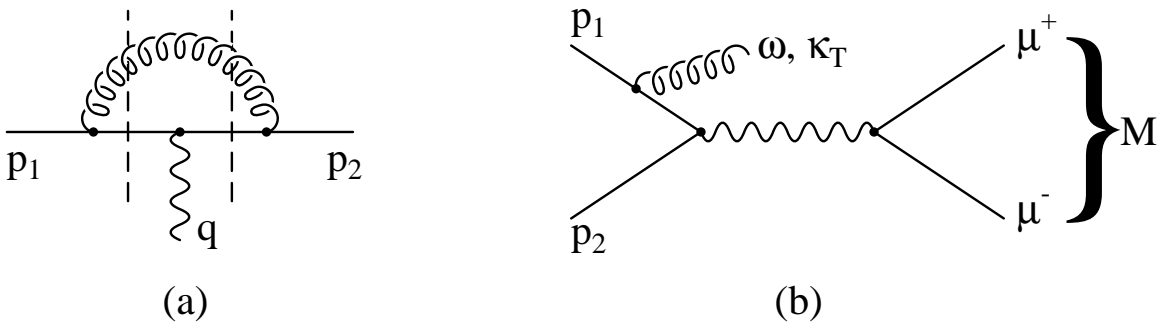}
\vskip -8mm
\caption[]{\label{fig6} {\em (a) Virtual and (b) real emission contributions 
to the Drell-Yan process $q (p_1) \overline{q} (p_2) \rightarrow \gamma^* 
(q)$.}} 
\end{center}
\end{figure}
Let us first recall the origin of the ${\cal O} (\pi^2 \alpha_S/\pi)$ term 
in the simplest, Drell-Yan, case, which arise from the diagrams of Fig.\ 6.  
We first consider the emission of a single 
soft gluon, shown in Fig.\ 6b, with energy $\omega$ and transverse momentum 
$\kappa_T$. 
The probability to emit the gluon is 
\begin{equation}
\overline{n} \; = \; 4 C_F \: \int^{\frac{1}{4} M^2} \: 
\frac{d\kappa_T^2}{\kappa_T^2} \: 
\int_{\kappa_T}^{\frac{1}{2} M} \: \frac{d\omega}{\omega} \: 
\frac{\alpha_S}{4 \pi} \; = \; \frac{\alpha_S C_F}{4 \pi} \: 
\ln^2 \: \frac{M^2}{\mu^2},
\label{eq:d50}
\end{equation}
where here $M$ is the invariant mass of the Drell-Yan $\mu^+\mu^-$ pair, and
where for simplicity of presentation we ignore, for the moment, the running 
of $\alpha_S (\kappa_T^2)$.  Such soft gluons are emitted independently.  
Summing over all possible emissions, we 
obtain a Poisson distribution with average multiplicity $\overline{n}$.  The 
normalization factor $\exp (- \overline{n})$ comes from the loop diagrams 
where the emitted gluon is absorbed by the other quark.  The 
factor is \cite{sud} 
\begin{equation}
\exp \: \left ( - \frac{\alpha_S C_F}{4 \pi} \; \ln \left ( \frac{M^2}{p^2_1} 
\right ) \; \ln  \left ( \frac{M^2}{p^2_2}\right ) \right )
\label{eq:e50}
\end{equation}
where each $\ln(M^2/p^2_i)$ contains $i\pi$ from the negative value of the
quark virtuality $p^2_i < 0$.\footnote{There are no $i\pi$ terms for the 
annihilation process $e^+ e^- \rightarrow q \overline{q}$ where 
the quark masses 
$p^2_i > 0$, or for elastic scattering where $M^2 = q^2 < 0$ and
$p^2_i - m^2 \leq 0$. The cut-off $\mu^2$ in (\ref{eq:d50}) is $\mu^2 \sim 
\vert p^2_i\vert$.}
Now on taking the sum of the virtual and real gluon emission 
terms we can obtain 
the ${\cal O} (\alpha_S)$ contribution to the cross section for Drell-Yan
production. We have
\begin{equation}
\sigma \; = \; \sigma_0 \: \left ( \left | 1 \: - \: 
\frac{\alpha_S C_F}{4 \pi} \: \ln^2 \: \left ( \frac{- M^2}{\mu^2} \right ) 
\right |^2 \: + \: 2 \: \frac{\alpha_S C_F}{4 \pi} \: \ln^2 \: 
\frac{M^2}{\mu^2} \right )
\label{eq:g50}
\end{equation}
where $\sigma_0$ is the lowest order cross section and $\mu^2 \sim \vert
p^2_i\vert$. 
The factor of 2 in the last term arises because the real emission can 
occur from either the $q$ or 
the $\overline{q}$ quark.  Thus, since $\ln (-M^2) = \ln M^2 + 
i \pi$, the \lq\lq real" and \lq\lq virtual" $\alpha_S \ln^2 M^2$ 
cancel and we have
\begin{eqnarray}
\sigma & = & \sigma_0 \: \left (1 + \frac{\alpha_S C_F}{2 \pi} \: \pi^2 
\: + \: \ldots \right ) \nonumber \\
& = & \sigma_0 \: \left (1 + {\textstyle \frac{1}{2}} \: 
\alpha_S C_F \pi \: + \: \ldots \right ) \nonumber \\
& \Rightarrow & \sigma_0 \: \exp \left ({\textstyle \frac{1}{2}} 
\: \alpha_S C_F \pi \right ),
\label{eq:f50}
\end{eqnarray}
where the last result corresponds to the resummation of the gluon emissions.  
If we, as we should, use running $\alpha_S (\kappa_T^2)$ in the 
integral of (\ref{eq:d50}), then we obtain a $\ln(M^2)\ln(\ln M^2)$ 
form.  Proceeding as before, and noting $\ln \ln (- M^2) = \ln \ln M^2 + 
(i\pi/\ln M^2)$, it is straightforward to show that the enhancement factor 
is such that the Drell-Yan cross section becomes
\begin{equation}
\sigma \; = \; \sigma_0 \: \exp (\alpha_S C_F \pi),
\label{eq:y1}
\end{equation}
with the argument of $\alpha_S$ equal to $M^2/4$.  Note the absence of the 
factor $\frac{1}{2}$ which occurs in the 
fixed $\alpha_S$ result (\ref{eq:f50}).  
The larger enhancement is expected since the running 
of $\alpha_S (\kappa_T^2)$ 
in (\ref{eq:d50}) weights the integrals towards smaller $\kappa_T$ values.

\begin{figure}[htb]
\begin{center}
\vskip -6mm
\leavevmode
\epsfxsize=16.cm
\epsffile[70 320 500 550]{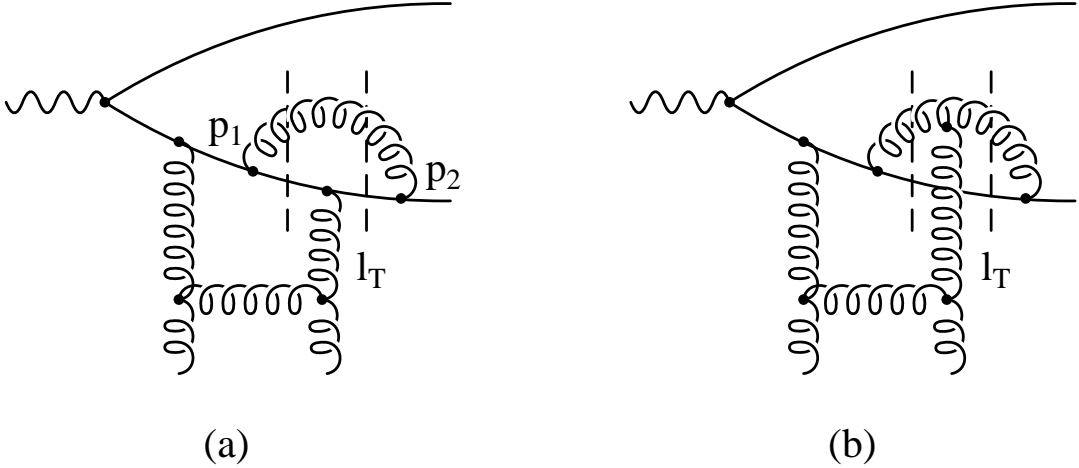}
\vskip -8mm
\caption[]{\label{fig7}{\em The (virtual) diagrams responsible for the $\pi^2$ 
enhancements to diffractive $c\overline{c}$ production.  The exchanged gluon 
has virtuality $\ell^2 < 0$.}} 
\end{center}
\end{figure}

The same $\pi^2$ enhancement occurs in diffractive open charm 
production\footnote{We are grateful to A. Kataev for discussions 
concerning the virtual
corrections.}.  It arises (in the Feynman gauge) from the product of the 
imaginary parts which correspond to the discontinuities indicated by 
the dashed lines in the diagrams shown in Fig. \ref{fig7}.  Here 
the exchanged gluon has virtuality $\ell^2 < 0$,
whereas the quarks have $p^2_i > 0$. Note that $\ell^2$ is the counterpart of 
$M^2$ for Drell-Yan production. These are the only graphs which have negative
values of the arguments ($\ell^2 / p^2_i < 0$) of both the logarithms of
the counterpart of (\ref{eq:e50}). Without the second $t$-channel 
gluon (indicated by $\ell_T$ in 
Fig. \ref{fig7}) the diagrams are analogous to ordinary 
deep inelastic scattering for which there are no 
large ${\cal O} (\alpha_S \pi)$ 
enhancements. Of course there may be other ${\cal O} (\alpha_S)$ contributions 
but these are expected to be much smaller than the $\pi^2$ terms. Thus a good 
estimate of the cross section for diffractive open charm production is
\begin{equation}
\sigma^{c\overline{c}} = \sigma_0^{c\overline{c}} \exp(\alpha_S C_F \pi),
\label{eq:y1a}
\end{equation}
as in Drell-Yan production, but in this case the $\pi^2$ enhancement arises
from quite different diagrams. For the diffractive process the argument of 
$\alpha_S$ is of the order of the largest mass squared of the 
quark-gluon state, that is
${\rm max} (p^2_1, p^2_2)$, which is of the order of $k^2_T + m^2$.  The 
largest value is $M^2$, where $M$ is the invariant mass 
of the $c\overline{c}$ system. Here we take $\alpha_S (M^2/4)$. 

The $\pi^2$ enhancement encapsulated in (\ref{eq:y1a}), which arises from the 
diagrams of Fig.\ 7, is not what we would naively expect.  In the space-time 
picture of section 2.1 we showed that the interaction time~$\tau_i$ of the 
$c\overline{c}$ pair with the proton is much less than 
the $\gamma^* \rightarrow 
c\overline{c}$ fluctuation time $\tau_\gamma$.  We may therefore hope that the 
corrections of Fig.\ 7 are suppressed by a factor $\tau_i/\tau_\gamma \ll 1$.  
This is indeed true for the contributions coming from the real part 
of the diagram, 
but it is not correct for those arising from the imaginary part, that is the 
crucial $i\pi$ terms.  The imaginary part (the discontinuity) of the amplitude 
corresponds to the possibility of producing a real (not virtual) intermediate 
state which may live infinitely long.  To obtain the imaginary contribution 
$i\pi \delta (p^2 - m_{qg}^2)$ we must 
integrate over the entire time interval from 
$\tau = - \infty$ up to $\tau = + \infty$ for the Feynman diagram 
in the coordinate 
representation.  When we evaluate the discontinuity 
using the unitarity relation, 2 Im$A = AA^*$, in the space-time picture 
the time in 
$A^*$ goes in the reverse direction to that in $A$.  From the formal point of 
view the inverse direction of time arises from the opposite signs of 
$i\varepsilon$ in the Feynman propagators 
$(p^2 - m_{qg}^2 \pm i\varepsilon)^{-1}$ of 
the amplitudes $A$ and $A^*$.  This point has been discussed in detail in the 
appendix of a paper by Gribov \cite{vng}.

Thus the large $K$ factor of $1 + {\cal O} (\pi^2 \alpha_S/\pi)$ does not 
contradict the space-time picture and the factorization properties described 
in section 2.1.  In other words this ${\cal O} (\pi \alpha_S)$ contribution 
may be regarded as what is left after the cancellation of the real and 
virtual gluon emission contributions which occur during the time-like 
evolution (or parton showering in the final state), which is usually 
not taken into account, as the leading logarithmic 
contributions for inclusive cross sections are cancelled.

It is interesting to contrast the enhancement of diffractive open charm 
production with the situation for diffractive $J/\psi$ production.  To 
calculate the probability for the exclusive $J/\psi$ process we have 
to convolute the $c\overline{c}$ amplitude ${\cal 
M}_{\lambda\lambda^\prime}$ of (\ref{eq:a8}) with the wave function of 
the $J/\psi$ meson just after the interaction.  Here there is not 
sufficient time and energy to form a real intermediate quark-gluon 
state (analogous to the states formed in Fig.\ 7).  Thus we do not expect 
a large $K$ factor for diffractive $J/\psi$ production.  The 
corrections coming from the loop diagrams may be treated as corrections 
to the $J/\psi$ wave function (including the possibility of 
$c\overline{c}g$ states in the $J/\psi$ meson).

We now return to diffractive open charm production, the subject of 
this paper.  We have considered the virtual 
corrections to $c\overline{c}$ production.  For 
completeness let us also consider the corrections 
for $c\overline{c}g$ production.  
In this case we are less able to estimate the enhancement factor.  In an 
important region of phase space the $(c\overline{c})g$ 
system is produced in a colour octet-octet configuration, with the 
$t$-channel gluon $\ell$ interacting with the $s$-channel gluon jet, as in
Fig.\ \ref{fig4}a.  Similar arguments lead to an
enhancement of the form of (\ref{eq:y1}) but with the colour factor $C_F$
replaced by $C_A$
\begin{equation}
\sigma^{c\overline{c}g} = \sigma_0^{c\overline{c}g} \exp (\alpha_S C_A \pi).
\label{eq:y2}
\end{equation}
On the other hand if the colour triplet-triplet 
configuration $c(\overline{c}g)$ 
is appropriate then the enhancement factor is that given in (\ref{eq:y1}).  
However, the $c\overline{c}g$ contribution is only appreciable at the large 
values of $M^2$, see section 4, and so the large value (and the 
large uncertainty) 
of the enhancement factor does not have serious impact on our main open charm 
predictions for HERA.  To illustrate the ambiguity we show results for two 
possible enhancements of the $c\overline{c}g$ contribution.  First, as 
an upper 
limit, we use (\ref{eq:y2}).  Then in a crude attempt to allow for the 
$c(\overline{c}g)$ and $\overline{c} (cg)$ configurations we take
\begin{equation}
\sigma^{c\overline{c}g}  \; = \; \sigma_0^{c\overline{c}g} \: 
\left [\frac{1}{3} \: \exp 
(\alpha_S C_A \pi) \: + \: \frac{2}{3} \: \exp (\alpha_S C_F \pi) \right ],
\label{eq:b64}
\end{equation}
which may be regarded as a lower limit.

The exponential enhancement factor 
for $c\overline{c}$ production in (\ref{eq:y1a}) 
is important.  It is typically in the range 2.7--4.0.  
However (\ref{eq:y1a}) is only an estimate 
of the true $K$ factor.  To remove the scale dependence we need to know  
the two loop diagrams.  Even at one loop level only the simplest, although 
the dominant, contribution (proportional to $\pi^2$) is taken into account.  
Nevertheless experience of the Drell-Yan process leads us to expect that the 
true $K$ factor is well approximated by the simple expression 
in (\ref{eq:y1a}). 

For diffractive $J/\psi$ production we do not have to consider the soft gluon 
emissions, since they are already included in the 
normalization of the $J/\psi$ 
wave function.  The ambiguity in the estimates of the relativistic corrections 
to the wave function \cite{rrml} is, in a sense, similar to the uncertainty 
due to the neglect of the higher loop etc.\ corrections to the simplified 
expression for the $K$ factor.

\bigskip
\noindent {\large \bf 4.  Numerical estimates of open charm
production}

We use the QCD formalism developed in the previous two sections to
estimate the rate of the diffractive production of open charm
from both longitudinally and transversely polarised photons.  We
discuss both photo- and electro-production.

\bigskip
\noindent {\bf 4.1.  A first glimpse of the structure of the $\gamma^*
\rightarrow c\overline{c}$ diffractive cross sections}

Before we present the detailed perturbative QCD predictions for the cross 
sections for diffractive open charm production it is informative to use the 
leading logarithmic approximation, (\ref{eq:d29}) and (\ref{eq:e29}), to 
crudely anticipate some of the general features of the results.  An 
important advantage of the
diffractive production of heavy, as opposed to light, quarks is
the dominance of small-distance contributions.  The $k_T$
integrations in (\ref{eq:d29}) and (\ref{eq:e29}) are infrared
safe, protected by the mass $m$ of the charm quark.  Moreover the
scale $K^2$ at which the gluon distribution is sampled is \cite{blw,gnz}
\begin{equation}
K^2 \; = \; \frac{m^2 + k_T^2}{1 - \beta}
\label{eq:b56}
\end{equation}
which may be considerably in excess of $m^2$ depending on the
value of $\beta$ and the dominant region of $k_T^2$ sampled. 
Fig.\ 8 shows the integrands $I_{L,T} (k_T^2)$ of the integrals
in (\ref{eq:d29}) and (\ref{eq:e29}) for various values of $Q^2$
and $\beta$ defined such that
\begin{equation}
x_{\funp} \: \frac{d\sigma_{L,T}}{dx_{\funp}} \; \equiv \;
\int_0^{\frac{1}{4} M^2 - m^2} \: \frac{dk_T^2}{\sqrt{1 -
4m_T^2/M^2}} \; I_{L,T} (k_T^2)
\label{eq:c56}
\end{equation}
where the cross sections have been integrated over $t$ assuming
the form $\exp (-b |t|)$ with slope\footnote{Elastic $J/\psi$ photoproduction 
at HERA is observed to have a 
slope $b = 4.0 \pm 0.2 \pm 0.2$ GeV$^{-2}$ \cite{h1b}.} 
$b = 4$ GeV$^{-2}$ \cite{jps}.  We comment on these plots below.

\begin{figure}[htb]
\begin{center}
\leavevmode
\epsfxsize=18.cm
\epsffile[50 310 500 550]{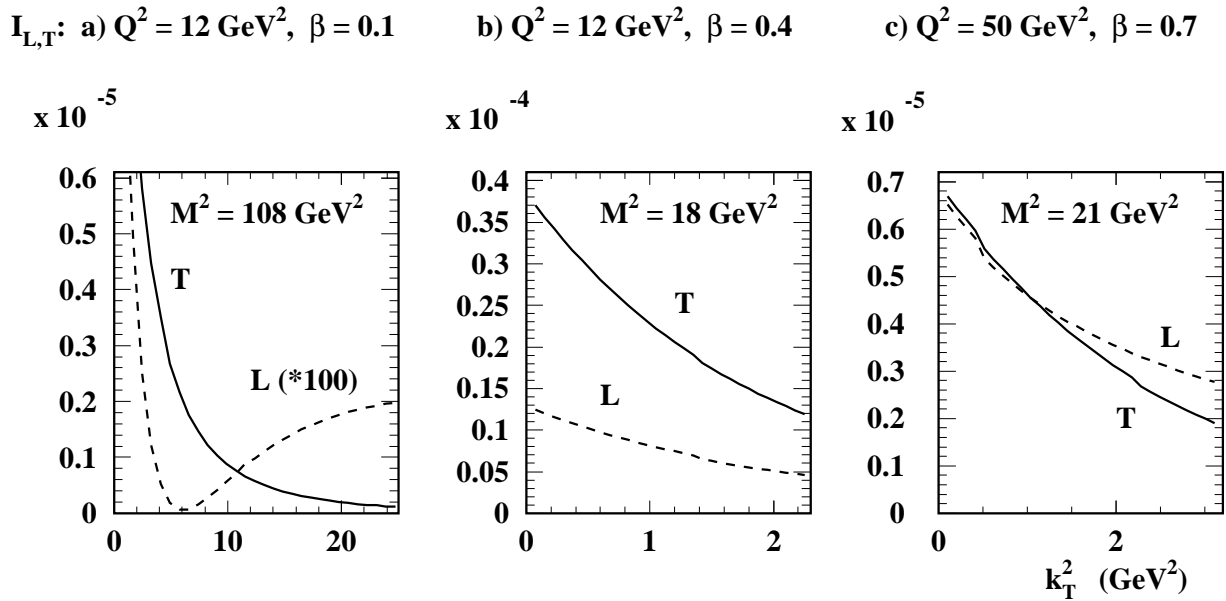}
\vskip -10mm
\caption[]{\label{fig8}{\em The $k_T^2$ integrands $I_{L,T} (k^2)$, in units
of GeV$^{-4}$ as defined in (\ref{eq:c56}), occurring in the
diffractive open charm cross section formulae, (\ref{eq:a28}) and
(\ref{eq:a29}), for $x_{\funp} = 5 \times 10^{-4}$ and various
values of $Q^2$ and $\beta$.  MRS(A$^\prime$) partons \cite{mrsa}
are used.}} 
\end{center}
\end{figure}

For moderate values of $\beta$ ($0.3 \lapproxeq \beta \lapproxeq
0.8$) we can obtain from the diffractive cross section formulae
in (\ref{eq:d29}) and (\ref{eq:e29}) a rough estimate of the
ratio $\sigma_L/\sigma_T$.  We find
\begin{eqnarray}
\sigma_L : \sigma_T & \approx & \frac{2 \langle m_T^2
\rangle \beta^3}{Q^4} \; : \; \frac{\beta (1 - \beta)^2}{Q^2} \nonumber
\\
& & \nonumber \\
& = & \frac{2 \langle m_T^2 \rangle Q^2}{(Q^2 + M^2)^3} \; : \;
\frac{M^4}{(Q^2 + M^2)^3} 
\label{eq:d56}
\end{eqnarray}
where $\langle m_T^2 \rangle$ is the average value of $m_T^2
\equiv m^2 + k_T^2$ sampled by the integration in (\ref{eq:d29}),
and where the factor of 2 in $\sigma_L$ comes from an approximate 
comparison of the integrands of (\ref{eq:d29}) and
(\ref{eq:e29}).  Eq.\ (\ref{eq:d56}) not only gives a reasonable
estimate of $\sigma_L/\sigma_T$ (up to a factor of two) but also
can be used as a rough guide of the
$Q^2$ and $M^2$ dependence of the individual cross sections ---
apart, of course, from the $M^2$ threshold effect coming from the
rapidly expanding kinematic region of integration $(0, \:
\frac{1}{4} M^2-m^2)$.  From (\ref{eq:d56}) we see
that for $\sigma_L$ to dominate over $\sigma_T$ we have to go to
large $Q^2$ and small $M^2$.  For example for $Q^2 = 50$ GeV$^2$
and $M^2 = 12$ GeV$^2$ we anticipate from (\ref{eq:d56}) that
$\sigma_L/\sigma_T \approx 4Q^2 m^2/M^4 \approx 3$ whereas if
$M^2$ is increased to 20 GeV$^2$ then $\sigma_L \approx
\sigma_T$.  Here we have used $\langle k_T^2 \rangle \approx
m^2$, see Fig.\ 8(c), and hence $\langle m_T^2 \rangle \approx
2m^2$.  These rough estimates of $\sigma_L/\sigma_T$ are in
agreement with the results of the full numerical evaluation of
the cross sections (which we will show in Fig.\ 10).

\begin{figure}[htb]
\begin{center}
\leavevmode
\epsfxsize=15.cm
\epsffile[50 250 530 580]{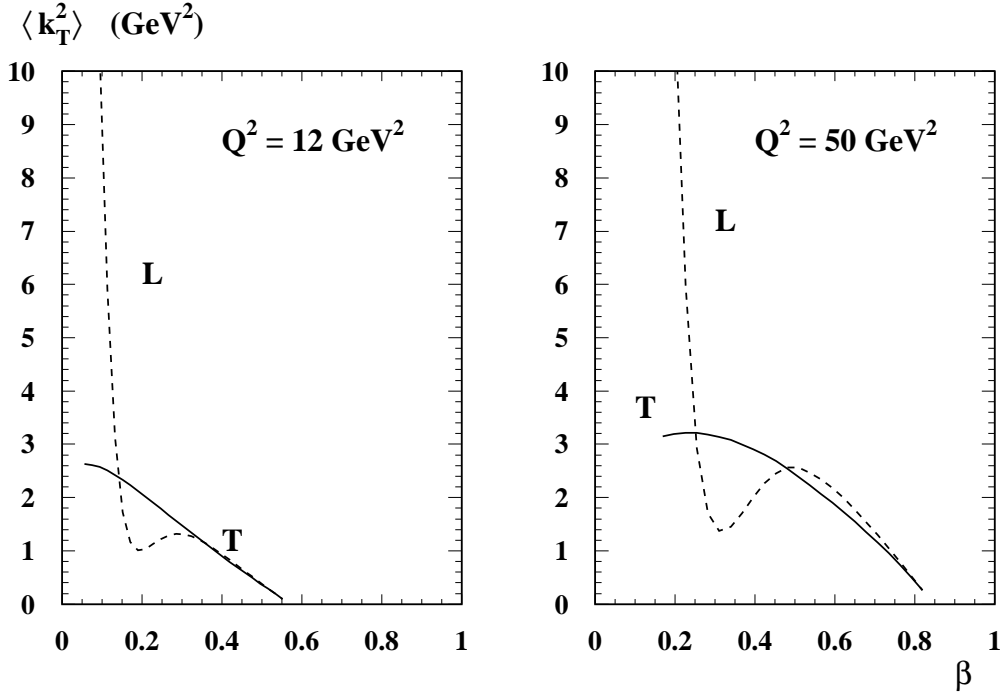}
\vskip -8mm
\caption[]{\label{fig9} {\em The average value of $\langle k_T^2 \rangle$ for
the integrals in (\ref{eq:a28}) and (\ref{eq:a29}) for $Q^2 = 12$
and $50$ GeV$\,^2$ and $x_{\funp} = 5 \times 10^{-4}$.  The
MRS(A$^\prime$)
gluon is used.}} 
\end{center}
\end{figure}
To see the typical values of $k_T^2$ sampled in the 
integrations (\ref{eq:d29}) 
and (\ref{eq:e29}) we calculate the average value of $\ln k_T^2$.  We show 
the results in Fig.\ 9 in the form
$$
\langle k_T^2 \rangle \; \equiv \; \exp \: \left (\overline{\ln k_T^2} \right )
$$
as a function of $\beta$ for our selected values of $Q^2$ and $x_\funp$.  
Unless $\beta$ is small, we see, for both $\sigma_T$ and
$\sigma_L$, that $\langle k_T^2 \rangle$ is of the order of
$m^2$; as may be expected, for example, from Figs.\ 8(b,c).  Of
course as we approach the kinematic bound at $\beta = Q^2/(Q^2 +
4m^2)$, which corresponds to $M^2 = 4m^2$, the interval of
integration shrinks to zero and hence $\langle k_T^2 \rangle
\rightarrow 0$.  On the other hand, if we
consider large $M^2$, so that we have a large interval for the
$k_T^2$ integration in (\ref{eq:d29}), then the integrand $I_L$
develops a different character.  It is dominated by contributions
near the upper end of the region of integration (see, for
example, the dashed curve in Fig.\ 8(a)).  It follows that the
value of $\langle k_T^2 \rangle$ is nearer the upper limit
$\frac{1}{4} M^2$ (as shown by the behaviour of the dashed curve
in Fig.\ 9 for small $\beta$).  Hence $\sigma_L$ samples the
gluon at high scales $K^2 \approx \frac{1}{4} M^2$. 
Unfortunately in this domain the cross section $\sigma_L$ is very
small.

\bigskip
\noindent {\bf 4.2.  Evaluation of the cross sections $\sigma^{L,T}$ for 
diffractive $c\overline{c}$ production}

We calculate the cross sections for the diffractive production of open charm 
from the QCD formula of (\ref{eq:a28}) and (\ref{eq:a29}) for $c\overline{c}$
production, together with the higher order contributions arising from 
$c\overline{c}g$ processes of section 3.1 and the $K$ factor of section 3.2.  
The integrals
\begin{equation}
{\cal I}_i (k_T^2) \; = \; \int_0^\infty \: \frac{d\ell_T^2}{\ell_T^2} \: 
\alpha_S (\ell_T^2) \: f (x_\funp, \ell_T^2) \: K_i (\ell_T^2, k_T^2)
\label{eq:z1}
\end{equation}
over the transverse momenta of the exchanged gluons are evaluated numerically, 
where \linebreak $\ell_T^2 K_i (\ell_T^2, k_T^2)$ are 
the expressions given in brackets in the formulae (\ref{eq:c24}) and 
(\ref{eq:c28}) for ${\cal I}_1$ and ${\cal I}_2$ respectively.  
The contributions from the infrared region $\ell_T^2 < Q_0^2$ are evaluated in 
two alternative ways.  We rewrite (\ref{eq:z1}) as
\begin{equation}
{\cal I}_i \; = \; \alpha_S (Q_0^2) \: A_i (Q_0^2) \: + \: \int_{Q_0^2}^\infty 
\: d\ell_T^2 \: \alpha_S (\ell_T^2) \: \frac{\partial \biggl (
x_\funp g (x_\funp, \ell_T^2) \biggr )}{\partial \ell_T^2} \: 
K_i (\ell_T^2, k_T^2),
\label{eq:z2}
\end{equation}
where we have used (\ref{eq:d17}) to express the unintegrated distribution in 
terms of the conventional gluon density.  The first estimate of the infrared 
contribution is made using
$$
A_i \; = \; x_\funp \: g (x_\funp, Q_0^2) \: K_i (0, k_T^2).
$$
We compare this result with that obtained by performing 
the $0 < \ell_T^2 < Q_0^2$ 
integration assuming that $\alpha_S (\ell_T^2) \: g (x_\funp, \ell_T^2) = 
(\ell_T^2/Q_0^2) \alpha_S (Q_0^2) \: g (x_\funp, 
Q_0^2)$.  The two methods give essentially the same results.  We also 
test the sensitivity 
of the predictions to variations in the choice of $Q_0^2$ in the range 0.65 to 
1.5 GeV$^2$.  Again we find that the results are stable.

So far we have calculated the imaginary part of the 
amplitude, see (\ref{eq:a9}).  
At high energy $W$, that is small $x_\funp$, our positive-signature exchange 
amplitude behaves as
\begin{equation}
T \: \propto \: i (x_\funp^{- \lambda} \: + \: (-x_\funp)^{- \lambda})
\label{eq:z3}
\end{equation}
arising from the small $x_\funp$ behaviour of the gluon.  Provided 
that $\lambda$ is small, $\lambda \lapproxeq 0.3$, the amplitude is 
dominantly imaginary and the 
real part can be calculated as a perturbation
\begin{equation}
\frac{{\rm Re} T}{{\rm Im} T} \; \approx \; \frac{\pi}{2} \lambda \; 
\approx \; 
\frac{\pi}{2} \:\frac{\partial 
\ln (x_\funp g (x_\funp, \ell_T^2))}{\partial \ln 
(1/x_\funp)}.
\label{eq:z4}
\end{equation}
We calculate the correction from (\ref{eq:z4}) and allow for a 
possible dependence of $\lambda$ on the scale $\ell_T^2$ of the 
gluon.  The contribution of the real 
part is included in all the predictions shown below.

\begin{figure}[htb]
\begin{center}
\vskip -10mm
\leavevmode
\epsfxsize=15.cm
\epsffile[30 90 520 760]{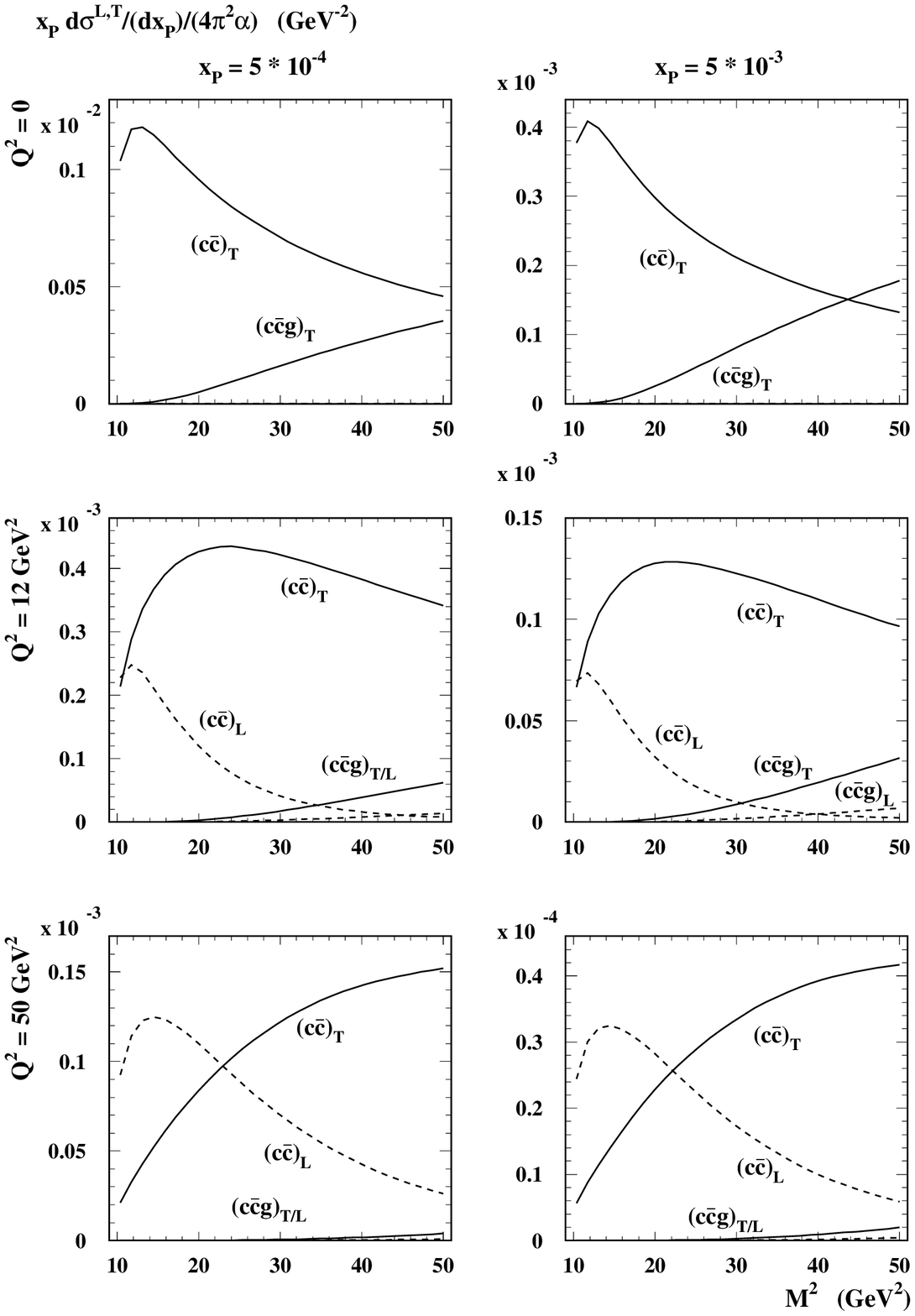}
\vskip -6mm
\caption[]{\label{fig10} {\em The cross sections $(x_{\funp}
d\sigma^{L,T}/dx_{\funp})/4 \pi^2 \alpha$ (in GeV$\,^{-2}$) for the
diffractive production of open charm from longitudinally and
transversely polarised photons for $x_\funp = 5 \times 10^{-4}$ and $5 \times 
10^{-3}$ and $Q^2 = 0, 12$ and $50$ GeV$\,^2$.  The continuous and dashed 
curves correspond to
$\sigma_T$ and $\sigma_L$ respectively.  Both the $c\overline{c}$
and $c\overline{c}g$ contributions are shown.  The mass of the
charm quark is taken to be $m = 1.5$ GeV. The MRS(A$^\prime$)
gluon is used.  The predictions are made using (\ref{eq:a28}), 
(\ref{eq:a29}) and (\ref{eq:a40}), and the \lq\lq $K$ factor" 
enhancements of (\ref{eq:y1a}), (\ref{eq:b64}) are not included.}} 
\end{center}
\end{figure}

The cross sections are shown in Fig.\ 10 in the form
\begin{equation}
\frac{x_\funp}{Q^2} F_i^{D (3)} \; \equiv \; \frac{1}{4 \pi^2 \alpha} \: 
x_\funp \: \frac{d \sigma^i}{dx_{\funp}} ,
\label{eq:a45}
\end{equation}
with $i = L,T$, after integration over $t$ assuming the form
$\exp (- b |t|)$ with the observed value of the slope parameter
$b = 4$ GeV$^{-2}$.  Eq.\ (\ref{eq:a45}) relates the cross sections that are 
shown in Fig.\ 10 to the conventional definition of $F^{D (3)}$.  The cross 
sections are plotted as a
function of the square of the invariant mass of the
$c\overline{c}$ system for six different values of $(x_{\funp},
Q^2)$;
namely $x_{\funp} = 5 \times 10^{-4}$ and $5 \times 10^{-3}$ and
$Q^2 =
0, 12$ and 50 GeV$^2$.  The gluon of the MRS(A$^\prime$) set of
partons \cite{mrsa} is used and the mass of the charm quark is
taken to be $m = 1.5$ GeV.  Throughout this
paper\footnote{Individual parton sets are, of course, evolved
with their respective QCD couplings.} we take the running
coupling $\alpha_S$ to be such that it gives $\alpha_S (M_Z^2) =
0.118$.  The uncertainty $\Delta \alpha_S (M_Z^2) = 0.005$
translates into typically a $\pm 20\%$ error on the predicted
cross sections.

Some of the predictions shown in
Fig.\ 10 are beyond the kinematic reach of the HERA electron-proton 
collider.  For example if we take the maximum $\gamma^* p$
energy for which open charm can be measured to be $W = 200$ GeV
then for $x_{\funp} = 5 \times 10^{-3}$ we see from (\ref{eq:a2})
that $Q^2 + M^2 < 200$ GeV$^2$, whereas for $x_{\funp} = 5 \times
10^{-4}$ we have $Q^2 + M^2 < 20$ GeV$^2$.

\begin{figure}[htb]
\begin{center}
\leavevmode
\epsfxsize=15.cm
\epsffile[70 210 500 640]{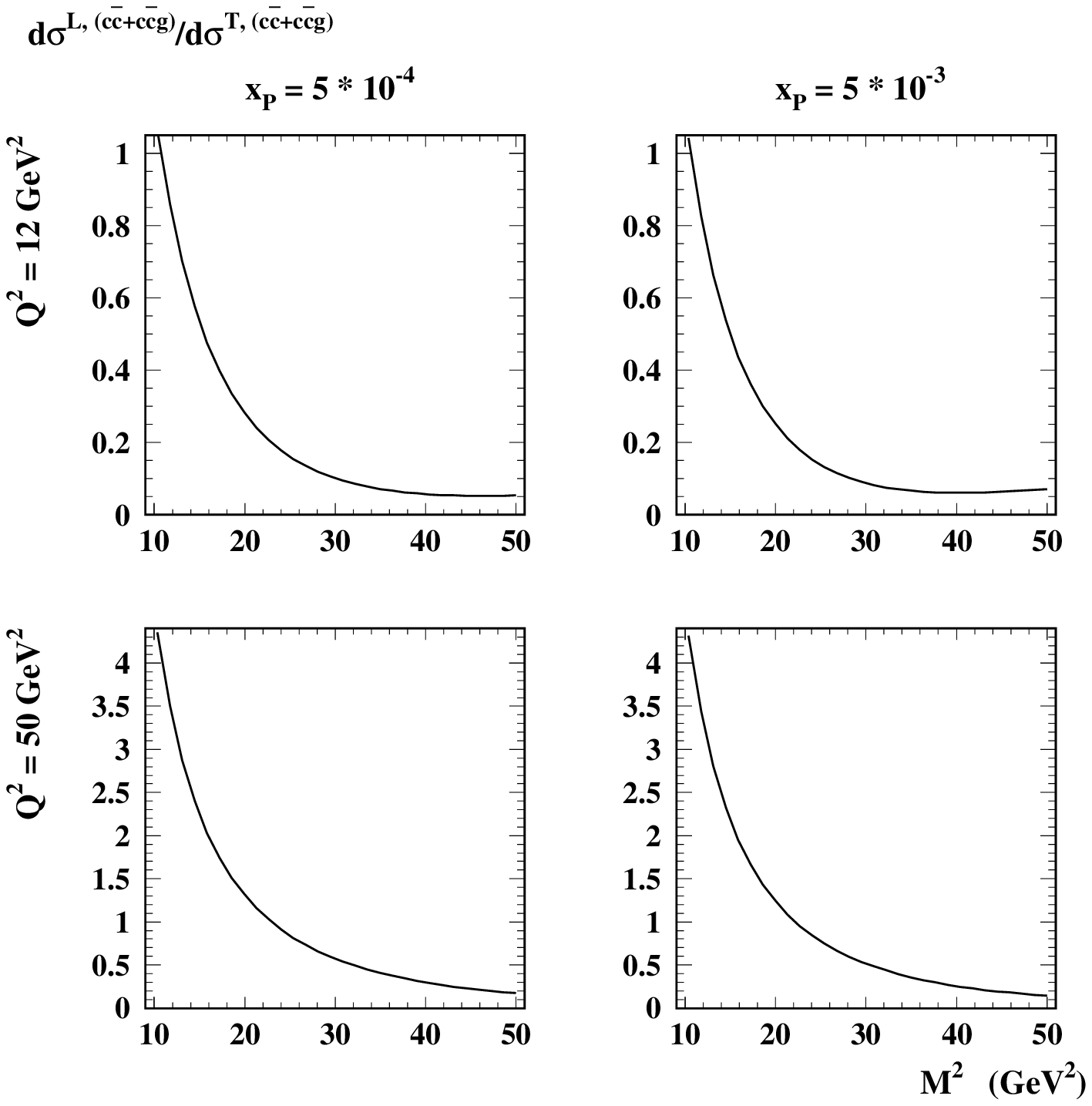}
\caption[]{\label{fig11} {\em The ratio $d\sigma^L/d\sigma^T$ obtained from
Fig.\ 10.}} 
\end{center}
\end{figure}

From Fig.\ 10 we see that the cross section $\sigma^T$ is
generally dominant, except for low values of $M^2$ at the higher
values of $Q^2$.  This property is clear from Fig.\ 11 which shows
the ratio $d\sigma^L/d\sigma^T$ as a function of $M^2$.  Indeed
the results for the $c\overline{c}$ contributions shown in Fig.\
10 quantify the structure that we anticipated in (\ref{eq:d56}). 
We notice that the gluon emission contributions $c\overline{c}g$
(of section 3.1) are relatively small for $M^2 \lapproxeq 25$
GeV$^2$, especially for $Q^2 \neq 0$.

\begin{figure}[htb]
\begin{center}
\leavevmode
\epsfxsize=15.cm
\epsffile[30 90 520 740]{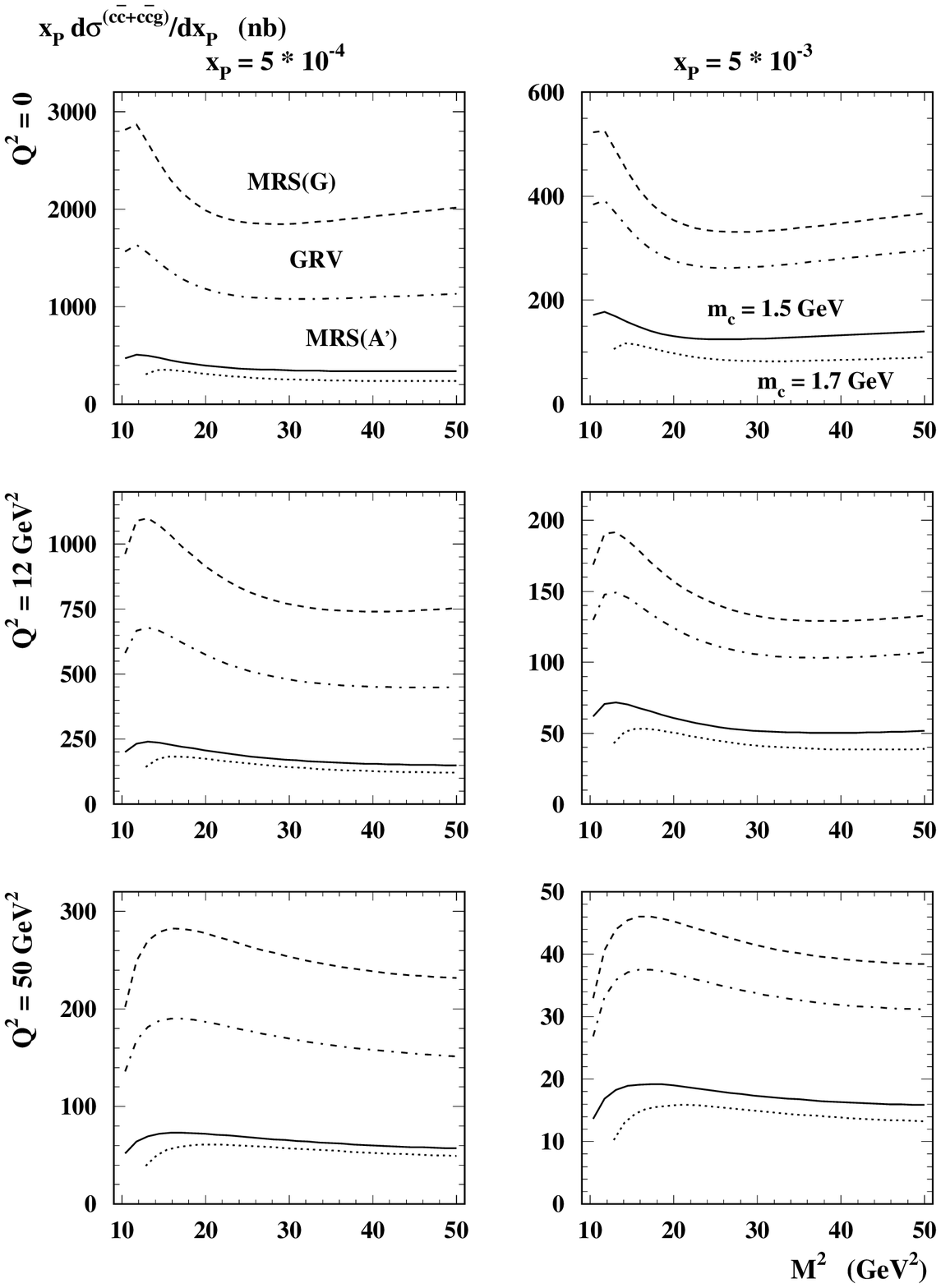}
\vskip -4mm
\caption[]{\label{fig12} {\em The cross section $(x_{\funp}
d\sigma/dx_{\funp}$ (in nb) for the diffractive
production of open charm obtained from the QCD formulae
(\ref{eq:a28}), (\ref{eq:a29}) and (\ref{eq:a40}) using
MRS(A$^\prime$,G) \cite{mrsa} (continuous, dashed) and GRV \cite{grv} 
(dot-dashed) partons (with $m
= 1.5$ GeV).  The dotted curve corresponds to a charm quark mass
$m = 1.7$ GeV for the MRS(A$^\prime$) gluon.  The \lq\lq $K$ factor" 
enhancements of (\ref{eq:y1a}) and (\ref{eq:b64}) are included.}} 
\end{center}
\end{figure}

Fig.\ 12 shows the sum of all the contributions to the cross
section for the diffractive production of open charm for three
recent sets of partons MRS(A$^\prime$,G) \cite{mrsa} and GRV
\cite{grv}.  The sensitivity to the square of the gluon density
is evident.    Moreover note that, as anticipated, the sensitivity 
is greater at the smaller value of $x_\funp$.  The most recent HERA data 
for the inclusive proton
structure function $F_2$ favour MRS(A$^\prime$) to GRV
but exclude MRS(G).  The predictions of Figs.\ 8--12 correspond
to the choice $m = 1.5$ GeV for the mass of the charm quark.  For
comparison the dotted curve in Fig.\ 12 shows the prediction for
$m = 1.7$ GeV for the MRS(A$^\prime$) gluon.

\clearpage

\bigskip
\noindent {\bf 4.3.  The $x_{\funp}$ dependence of the cross
sections}

For diffractive production it is conventional to plot the cross
section (integrated over $t$) versus $x_{\funp}$ on a $\log-\log$
plot
in order to see if there is a universal $x_{\funp}^{-n}$
dependence
with $n$ independent of $\beta$ and $Q^2$.  The motivation for
studying such a power-like dependence on $x_{\funp}$ comes from
the
Regge inspired approach to diffractive scattering in which $n$ is
closely related to the $(t = 0)$ intercept of the Pomeron.

\begin{figure}[htb]
\begin{center}
\leavevmode
\epsfxsize=15.cm
\epsffile[20 360 590 610]{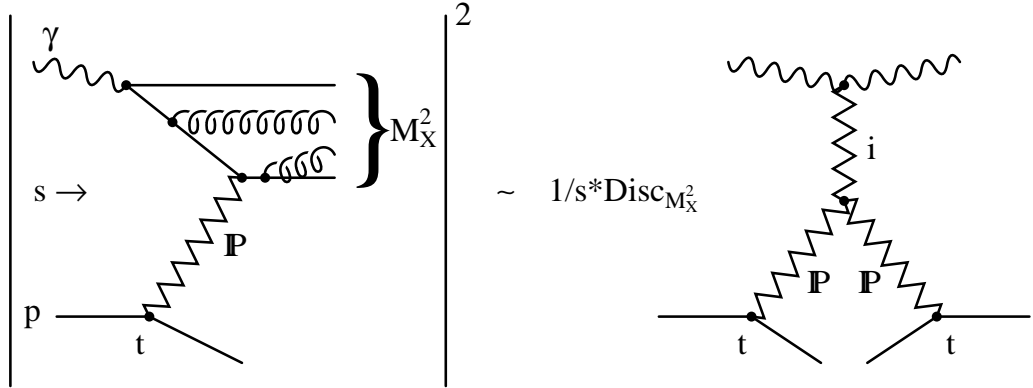}
\vskip -6mm
\caption[]{\label{fig13} {\em The diffractive cross section for $\gamma^* p
\rightarrow Xp$ for large $s/M_X^2$ and large $M_X^2$ expressed
in triple Regge form, where Disc$_{M_X^2}$ means the
discontinuity is to be taken across the $M_X^2$ cut of the elastic
photon-Pomeron amplitude.}} 
\end{center}
\end{figure}

Suppose we were to assume that at high energies our diffractive
process is dominated by Pomeron exchange.  Then it follows (see
Fig.\ 13) that the cross section for large $s/M_X^2$ may be
written in the form
\begin{equation}
\frac{d\sigma}{d x_{\funp} dt} \; \sim \; \sum_i \: G_{PPi} (t)
\:
\left ( \frac{s}{M_X^2} \right )^{2 \alpha_P (t) - 1} \:
(M_X^2)^{\alpha_i (0) - 1}
\label{eq:a47}
\end{equation}
where the sum is over all Regge exchanges which contribute to
photon-Pomeron elastic scattering including the Pomeron itself. 
This result applies for large $M_X^2$, but we can reasonably
expect the same form with $M_X$ replaced by $M$, for large values
of the $c\overline{c}$ invariant mass.  Recalling that
\begin{equation}
\frac{s}{M^2} \; \sim \; \frac{1}{x_{\funp}}, \hspace*{1cm} M^2
\;
= \;
\frac{Q^2}{\beta} \: (1 - \beta)
\label{eq:a48}
\end{equation}
we see that we can rewrite the cross section in the form
\begin{equation}
\frac{d\sigma}{d x_{\funp} dt} \; \sim \; F_2^P (\beta, Q^2, t)
\:
\left ( \frac{1}{x_{\funp}} \right )^{2 \alpha_P (t) - 1}
\label{eq:a49}
\end{equation}
where $F_2^P$ can be regarded as the structure function of the
Pomeron.  The main feature of this Pomeron exchange model is that
the power $n = 2 \alpha_P (t) - 1$ does not depend on $\beta$ and
$Q^2$ and so can be treated as a flux factor \cite{is}.

It is therefore informative to display our predictions for the
diffractive open charm cross sections
\begin{equation}
F_{L,T}^{D (3)} \: (x_{\funp}, \beta, Q^2) \; \equiv \; 
\frac{Q^2}{4\pi^2 \alpha} 
\: \int \: dt \: \frac{d\sigma^{L,T}}{d x_{\funp} dt}
\label{eq:a50}
\end{equation}
on a $\ln F^{D (3)}$ versus $\ln x_{\funp}$ plot to check whether
a
linear $x_{\funp}^{-n}$ form is obtained.  The results are shown
in
Fig.\ 14 for the values of $\beta$ and $Q^2$ that were chosen for
Fig.\ 8.  From the figure we see
that QCD predicts that the behaviour is only approximately 
linear.  The values of 
$n$ obtained from the slope of the curves at selected values of $x_\funp$ 
are shown in Table 1.  We see that $n$ depends on $Q^2, \beta$ and also 
$x_\funp$.  This is to be expected.  For the diffractive production of 
heavy quarks we 
can trust the purely perturbative QCD calculation based on
two-gluon exchange; we are not in the regime of (non-perturbative
or \lq\lq soft") Pomeron exchange where $n = 2 \alpha_P (\overline{t}) - 1$.

\begin{table}[htb]
\caption{The exponents $n$ of the effective $x_\funp^{-n}$ behaviour of 
$F^{D(3)}$ at different values of $x_\funp, Q^2$ and $\beta$. $Q^2$ is given 
in GeV$^2$.}
\begin{center}
\begin{tabular}{|c|cc|cc|cc|} \hline
& $Q^2 = 12$, & $\beta = 0.1$ & $Q^2 = 12$, & $\beta = 0.4$ 
& $Q^2 = 50$, & $\beta = 0.7$ \\ \cline{2-7}
\raisebox{1.5ex}[0pt]{$x_\funp$} & $n_T$ & $n_L$ & $n_T$ 
& $n_L$ & $n_T$ & $n_L$ \\ \hline
$3 \times 10^{-4}$ & 1.40 & 1.31 & 1.48 & 1.52 & 1.53 & 1.56 \\
$10^{-3}$ & 1.38 & 1.31 & 1.48 & 1.53 & 1.52 & 1.56 \\
$3 \times 10^{-3}$ & 1.38 & 1.31 & 1.53 & 1.59 & 1.59 & 1.63 \\
$10^{-2}$ & 1.43 & 1.39 & 1.54 & 1.59 & 1.59 & 1.63 \\ \hline
\end{tabular}
\end{center}
\end{table}

\begin{figure}[htb]
\begin{center}
\leavevmode
\epsfxsize=18.cm
\epsffile[50 320 500 520]{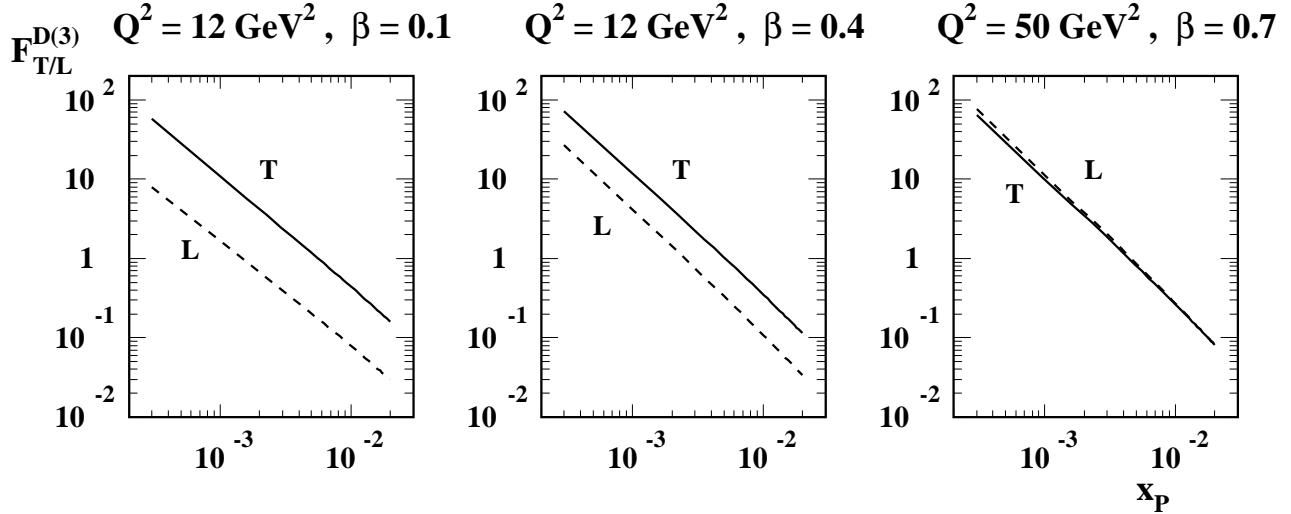}
\vskip -6mm
\caption[]{\label{fig14} {\em Plots of $F^{D (3)} (x_{\funp}, \beta, Q^2)$,
defined
in (\ref{eq:a50}), versus $x_{\funp}$ for the values of $\beta$
and
$Q^2$ used in Fig.\ 8.  As usual the continuous (dashed) curves
correspond to $F_T (F_L)$.  The values of the exponent $n$ of
$F^{D (3)} \sim x_{\funp}^{-n}$ are listed in Table 1.}} 
\end{center}
\end{figure}

\bigskip
\noindent {\bf 4.4.  The diffractive structure function $F_2^{D (2)}$}(charm)

It is also useful to introduce the diffractive structure function
integrated over $x_{\funp}$ (as well as $t$)
\begin{equation}
F_2^{D (2)} \: (\beta, Q^2) \; \equiv \; \frac{Q^2}{4 \pi^2
\alpha} \:
\int_{x_1}^{x_2} \: \frac{d\sigma}{d x_{\funp}} \:
dx_{\funp}
\label{eq:a51}
\end{equation}
which would be the charm component of the Pomeron structure
function in the Pomeron exchange model approach.  Even though we
have seen that there is no basis for this model, it is still
possible to write an evolution equation, similar to GLAP
evolution, which gives the $\beta$ and $Q^2$ dependence of
$F_2^{D (2)}$ \cite{lw,gs}.  Note that the integral over $\beta$
of $F_2^{D (2)}$ gives the contribution of diffractive charm
production to the total deep inelastic cross section and is
intimately related to the shadowing corrections to the deep
inelastic structure function.

\begin{figure}[htb]
\begin{center}
\leavevmode
\epsfxsize=10.cm
\epsffile[180 300 400 550]{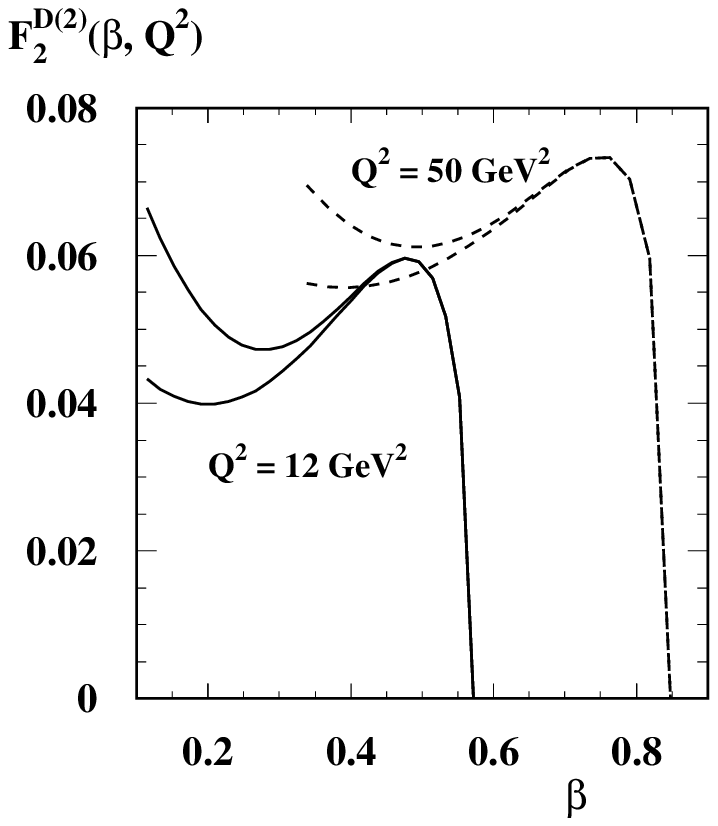}
\caption[]{\label{fig15} {\em The continuous and broken 
curves are the predictions 
for the contribution of diffractive open charm
production to the structure function $F_2^{D (2)} (\beta, Q^2)$
of (\ref{eq:a51}) as a function of $\beta$ for $Q^2 = 12$ and $50$ GeV$\,^2$ 
respectively.  The MRS(A$^\prime$) gluon is used.  $x_{\funp}$ is integrated 
over the interval ($0.0003$, $0.05$).  The curves 
branch at low values of $\beta$ 
according to whether we use the enhancement 
factor (\ref{eq:y2}) or (\ref{eq:b64}) 
arising from the virtual corrections to $c\overline{c}g$ production.}} 
\end{center}
\end{figure}

In Fig.\ 15 we show the predictions for
$F_2^{D (2)}$ as a function of $\beta$ for $Q^2 = 12$ and 50~GeV$^2$.  
We take the limits of the integration in (\ref{eq:a51}) to be 
$x_1 = 0.0003$ and $x_2 = 0.05$ so as to be able to compare our predictions 
for $F_2^{D (2)}$(charm) with the experimental measurements \cite{h1diff} 
for the total diffractive structure function (which includes light quark 
production as well as charm).  The experimental data at $Q^2 = 12$ GeV$^2$ 
give values of $F_2^{D(2)}$(total) in the range 0.15--0.2, approximately 
independent of $\beta$.  At $Q^2 = 50$ GeV$^2$ the measured values are 
\linebreak $F_2^{D(2)}$(total) $\simeq 0.3$ (0.2) 
at $\beta = 0.2$ (0.65), but again compatible within 
the errors with no $\beta$ dependence.  These are the values obtained 
by the H1 collaboration \cite{h1diff}.  Similar results are found 
by the ZEUS collaboration \cite{zeusdiff}.  We thus see from 
Fig.\ 15 that in this kinematic range the open charm contribution 
is predicted to be approximately 25--30\% of the measured 
diffractive structure function.  The results shown in Fig.\ 15 are obtained 
using MRS(A$^\prime$) partons and the mass of the charm quark $m = 1.5$ GeV.  
If a mass $m = 1.7$ GeV were to be used then the height of the peaks shown in 
Fig.\ 15 would be reduced by 20\%.  If on the other hand GRV or MRS(G) partons 
are used then the peak value of $F_2^{D(2)}$(charm) at $Q^2 = 50$ GeV$^2$ is 
found to be 0.17 or 0.24 respectively, as compared 
to 0.074 for MRS(A$^\prime$).  
When compared with the experimental values of $F_2^{D(2)}$(total), these high 
values of $F_2^{D(2)}$(charm) clearly disfavour 
the small $x$ gluon distributions of these sets of partons, even allowing for 
the uncertainty in the $K$ factor enhancement (\ref{eq:y1a}) or due to 
the choice of $m$.

Fig.\ 15 displays clearly the characteristic $\beta$ dependence of diffractive 
open charm production.  First we see the kinematic bound for $c\overline{c}$ 
production.  It corresponds to an upper limit on $\beta$,
\begin{equation}
\beta_{\rm max} \; = \; \frac{Q^2}{Q^2 + M_{th}^2}
\label{eq:a78}
\end{equation}
where the threshold mass is given by $M_{th}^2 = 4 m^2$.  The peak seen just 
below this value of $\beta$ arises from $c\overline{c}$ production just above 
threshold with a fall off as $M^2$ increases (i.e.\ $\beta$ decreases). 
The rise 
which occurs for smaller $\beta$ corresponds to $c\overline{c}g$ production.  
Recall that for the $c\overline{c}g$ component there is a large enhancement 
arising from virtual corrections to this process.  The size of the enhancement 
is not well known.  This uncertainty in the $c\overline{c}g$ contribution is 
represented in Fig.\ 15 by two curves which correspond to two 
different choices, 
(\ref{eq:y2}) and (\ref{eq:b64}), of the $c\overline{c}g$ enhancement $K$ 
factor.  Indeed at the lowest value of $\beta$ shown (for each $Q^2$), after 
the $c\overline{c}$ contribution is subtracted\footnote{The $c\overline{c}$ 
contribution vanishes linearly with $\beta$ as $\beta \rightarrow 0$.}, we 
find that the residual 
$c\overline{c}g$ contribution has approximately a factor of two uncertainty. \\

\newpage

\bigskip
\noindent {\large \bf 5.  Conclusions}

We have presented the QCD predictions for the cross sections
$(\sigma_{L,T})$ for the diffractive production of open charm from 
longitudinally 
and transversely polarised photons of virtuality $Q^2$.  At lowest order, the 
diffractive processes $\gamma^* p \rightarrow c\overline{c}p$ are driven by 
two-gluon exchange between the $c\overline{c}$ pair and the proton.  The 
perturbative predictions are protected by the mass $m$ of the charm quark and 
are infrared safe.  In fact the diffractive production of open charm depends 
on the square of the gluon density $x_\funp g (x_\funp, K^2)$, with $x_\funp = 
(Q^2 + M^2)/s$ at a scale
\begin{equation}
K^2 \; = \; \left ( \langle k_T^2 \rangle \: + \: m^2 \right ) \;
\left (1 \: + \: \frac{Q^2}{M^2} \right )
\label{eq:a52}
\end{equation}
where typically $\langle k_T^2 \rangle \sim m^2$.  $M$ is the invariant mass 
of the $c\overline{c}$ system.

We have aimed to make our diffractive cross section predictions as 
realistic as 
possible for the experiments at HERA.  We have studied the $Q^2$ and $M^2$ 
dependence of both $\sigma_L$ and $\sigma_T$.  We found that $\sigma_L$ only 
exceeds $\sigma_T$ in the kinematic region of low $M^2$ and high $Q^2$.  There 
are several new features incorporated in our analysis.  First, we integrate 
explicitly over the transverse momenta $\pm \mbox{\boldmath $\ell$}_T$ of the 
exchanged gluons.  This proves to be important, since we find 
that the inclusion 
of the $\ell_T^2$ effects increases the cross sections by about a 
factor of 2 as 
compared to using just the leading logarithmic approximation.  Second, we have 
estimated the higher-order contributions to diffractive open 
charm production.  
These were divided into the computation of real emission $c\overline{c}g$ 
contributions, and the estimation of the enhancement due to 
virtual corrections 
to $c\overline{c}$ production.  The real gluon emission 
contributions were found 
to be small for low $M^2$, $M^2 \lapproxeq 25$ GeV$^2$, and moreover 
decrease with 
increasing $Q^2$.  The virtual corrections on the other hand, are surprisingly 
important.  They arise from the diagrams of Fig.\ 7, and have some 
similarities 
to the $\pi^2$ enhancement of the ${\cal O} (\alpha_S)$ corrections that is 
well-known in Drell-Yan production.  However, their occurrence in diffractive 
$\gamma^* \rightarrow c\overline{c}$ production is a novel and theoretically 
interesting effect.  To a good approximation they can be resummed and 
represented by a $K$ factor 
enhancement in the form of an exponential with 
argument ${\cal O} (\alpha_S \pi)$.  
Just as in Drell-Yan production, we estimate an enhancement 
of the lowest-order 
result by a factor of about 3.  Due to the larger colour coupling 
of the gluon, 
the enhancement is expected to be greater 
for the $c\overline{c}g$ contribution.  
However, in this case the $K$ factor is not well-known.  Thus in 
the large $M^2$ 
(and small $Q^2$) region, where the $c\overline{c}g$ contribution 
is not negligible, 
our cross section predictions have a larger uncertainty (as represented, for 
example, by the branching of the curves shown in Fig.\ 15 at 
the lower value of $\beta$).

We have used the above perturbative QCD formalism to calculate the 
cross sections 
$\sigma_{L,T}$ for diffractive open charm production.  We have presented 
representative numerical results to illustrate the $Q^2$, $M^2$ and $x_\funp$ 
dependence of the cross sections, which are relevant to the 
experiments at HERA.  
In particular we show the sensitivity to the choice of gluon distribution at 
small $x$.

We may compare diffractive open charm production with diffractive $J/\psi$ 
production at HERA.  Both processes are special in that they depend 
quadratically on the gluon distribution at small $x$.  The main uncertainty 
in the calculation of the cross section for diffractive $J/\psi$ production 
was found to be associated with the Fermi motion of the $c$ and $\overline{c}$ 
within the $J/\psi$ and the choice of the mass $m$ of the charm quark 
\cite{rrml,fks}.  The uncertainty is greater in the predicted size of the 
cross section than in the energy (or equivalently the $x$) dependence of the 
cross section.  That is the shape, rather than the normalization, of 
the observed 
cross section is a better discriminator between the various gluon 
distributions.  
The shape of the diffractive $J/\psi$ photoproduction data collected at HERA 
favours the MRS(A$^\prime$) gluon relative to that of GRV, and rules out the 
MRS(G) gluon \cite{rrml}.

Diffractive open charm production has the advantage that it is independent of 
the uncertainties due to the $J/\psi$ wave function.  On the other 
hand the higher 
order corrections to open charm 
have (novel) $\pi^2$ enhancements which lead to 
a large $K$ factor which, at present, can only be estimated.  The $K$ factor 
uncertainty is not expected to be present in the $J/\psi$ prediction; 
it is automatically subsummed by using the observed leptonic width to fix the 
$J/\psi$ coupling.  High energy data for the two processes will therefore act 
as complementary and independent probes of the gluon at small $x$.  Their 
quadratic sensitivity to the gluon means that valuable information can already 
be obtained despite the above uncertainties.

When we compared our perturbative QCD predictions for diffractive open charm 
production with the inclusive data for diffractive production, at a given 
$\beta$ and $Q^2$, we estimate that 
about 25--30\% of diffractive events arise from 
$c\overline{c}$ production (if the MRS(A$^\prime$) gluon 
is used).  The $\beta$ 
(and $Q^2$) dependence of diffractive $c\overline{c}$ production has special 
features that are well-illustrated in the sample results shown in Fig.\ 15.  
Low mass $c\overline{c}$ production leads to a characteristic peak in 
the region $\beta 
\lapproxeq Q^2/(Q^2 + 4m^2)$, while $c\overline{c}g$ production only becomes 
important at much lower $\beta$.  To identify the dramatic threshold behaviour 
it will be necessary to reconstruct the mass of the $c\overline{c}$ system.

In summary we have used perturbative QCD to predict diffractive open charm 
production at HERA as a function of $Q^2$, $M^2$ and $x_\funp$.  The main 
unknowns are (i) the gluon distribution at small $x_\funp$, (ii) the mass $m$ 
of the charm quark and (iii) the accuracy of the estimate of the 
large $K$ factor 
enhancement.  The quadratic sensitivity to (i) means that information 
on the gluon 
can still be obtained despite the ambiguities arising 
from (ii) and (iii).  If 
the gluon is determined at small $x$ by independent means, then 
the characteristic 
$Q^2$, $M^2$ and $x_\funp$ dependence of 
diffractive $c\overline{c}$ production 
will offer a valuable probe of the validity of perturbative QCD at $x_\funp 
\lapproxeq 10^{-3}$ and at scales of the order of a few times $m^2$. \\ 

\bigskip
\noindent {\large \bf Acknowledgements}

We thank R.\ G.\ Roberts, A.\ Kataev, V.\ A.\ Khoze, 
P.\ J.\ Sutton and T.\ K.\ 
Gehrmann for useful discussions.  We thank the UK Particle Physics 
and Astronomy Research Council and the Royal Society for support, and Grey 
College of the University of Durham for their warm hospitality.  This
research was also supported in part (EML) by CNPq of Brazil and in 
part (MGR) by the Russian Fund of Fundamental Research 96 02 17994.


\newpage

\begin{thebibliography}{xx}
\bibitem{jps} H1 collaboration:  S.\ Aid et al., DESY preprint
96-037; \\
ZEUS collaboration:  M.\ Derrick et al., Phys.\ Lett.\ {\bf B350}
(1995) 120.

\bibitem{r} M.\ G.\ Ryskin, Z.\ Phys.\ {\bf C57} (1993) 89.

\bibitem{bfgms} S.\ Brodsky et al., Phys.\ Rev.\ {\bf D50} (1994)
3134.

\bibitem{rrml} M.\ G.\ Ryskin, R.\ G.\ Roberts, A.\ D.\ Martin
and E.\ M.\ Levin, Durham preprint DTP/95/96.

\bibitem{fks} L.\ Frankfurt, W.\ Koepf and M.\ Strikman, Tel-Aviv
University preprint, TAUP-2290-95.

\bibitem{bl} S.\ Brodsky and P.\ Lepage, Phys.\ Rev.\ {\bf D22}
(1980) 2157.

\bibitem{m} A.\ H.\ Mueller, Nucl.\ Phys.\ {\bf B335} (1990) 115.

\bibitem{nz} N.\ N.\ Nikolaev and B.\ G.\ Zakharov, Z.\ Phys.\
{\bf C53} (1992) 331.

\bibitem{blw} J.\ Bartels, H.\ Lotter and M.\ W\"{u}sthoff, DESY 
preprint 96-026.

\bibitem{gnz} M.\ Genovese, N.\ N.\ Nikolaev and B.\ G.\ Zakharov, 
hep-ph/9603285.

\bibitem{nz2} N.\ N.\ Nikolaev and B.\ G.\ Zakharov, Z.\ Phys.\ {\bf C49} 
(1991) 607; \\ Phys.\ Lett.\ {\bf B260} (1991) 414.

\bibitem{lw} E.\ M.\ Levin and M.\ W\"{u}sthoff, Phys.\ Rev.\
{\bf D50} (1994) 4306.

\bibitem{gr} M.\ Gl\"{u}ck and E.\ Reya, Phys.\ Lett.\ {\bf B83}
(1979) 98.

\bibitem{w} E.\ Witten, Nucl.\ Phys.\ {\bf B104} (1976) 445.

\bibitem{grs} M.\ Gl\"{u}ck, E.\ Reya and M.\ Stratmann, Nucl.\
Phys.\ {\bf B422} (1994) 37.

\bibitem{dyk} J.\ Kubar-Andre and F.\ E.\ Paige, Phys.\ Rev.\ {\bf D19} 
(1979) 221; \\
G.\ Altarelli, R.\ K.\ Ellis and G.\ Martinelli, Nucl.\ Phys.\ {\bf B143} 
(1978) 521; \\
ibid. {\bf B157} (1979) 461.

\bibitem{sud} V.\ V.\ Sudakov, JETP {\bf 3} (1956) 65.

\bibitem{vng} V.\ N.\ Gribov, Sov.\ Phys.\ JETP {\bf 57} (1970) 709.

\bibitem{mrsa} A.\ D.\ Martin, R.\ G.\ Roberts and W.\ J.\
Stirling, Phys.\ Lett.\ {\bf B354} (1995) 155.


\bibitem{h1b} H1 collaboration:  S.\ Aid et al., DESY preprint 96-037.

\bibitem{grv} M.\ Gl\"{u}ck, E.\ Reya and A.\ Vogt, Z.\ Phys.\
{\bf C67} (1995) 433.

\bibitem{is} G.\ Ingelman and P.\ Schlein, Phys.\ Lett.\ {\bf
B152} (1985) 256.

\bibitem{gs} T.\ Gehrmann and W.\ J.\ Stirling, Z.\ Phys.\ {\bf
C70} (1996) 89.

\bibitem{h1diff} H1 collaboration:  T.\ Ahmed et al., Phys.\ Lett.\ 
{\bf B348} (1995) 681.

\bibitem{zeusdiff} ZEUS collaboration:  M.\ Derrick et al., Z.\ Phys.\ 
{\bf C68} (1995) 569.
\end{thebibliography}
\end{document}